\documentclass[10pt,letterpaper]{article}
\usepackage{opex3}

\usepackage{graphicx}
\usepackage{hyperref}
\usepackage{cite}
\usepackage{color}
\usepackage[normalem]{ulem}
\usepackage[abs]{overpic}

\newcommand{\doi}[2]{\href{http://dx.doi.org/#1}{#2}}

\begin{document}

\title{
Phase-space properties of magneto-optical traps utilising micro-fabricated gratings.}
\author{J.\ P.\ McGilligan, P.\ F.\ Griffin,  E.\ Riis, and A.\ S.\ Arnold$^*$\email{*aidan.arnold@strath.ac.uk}}
\address{Dept.\ of Physics, SUPA,  University of Strathclyde, Glasgow G4 0NG, UK}


\begin{abstract}
We have used diffraction gratings to simplify the fabrication, and dramatically increase the atomic collection efficiency, of magneto-optical traps using micro-fabricated optics. The atom number enhancement was mainly due to the increased beam capture volume, afforded by the large area ($4\,$cm$^2$) shallow etch ($\sim 200\,$nm) binary grating chips. Here we 
provide a detailed theoretical and experimental investigation of the on-chip magneto-optical trap temperature and density in four different chip geometries using $^{87}$Rb, whilst studying effects due to MOT radiation pressure imbalance. With optimal initial MOTs on two of the chips we obtain both large atom number $(2\times 10^7$) \textit{and} sub-Doppler temperatures $(50\,\mu$K) after optical molasses.
\end{abstract}



\section{Introduction}

Laser cooling \cite{lascool} revolutionised atomic physics by making it possible to rapidly and robustly chill dilute clouds of atoms to below a millionth of room temperature. At such slow speeds it is possible to observe and interact with atoms a thousand times longer. This makes measurements of atomic transitions correspondingly more accurate, suitable for advanced quantum metrological devices. The starting point for the majority of cold atom experiments is a magneto-optical trap (MOT) \cite{raab87,monroe}. A MOT both cools and traps atoms surrounded in red-detuned laser light via changes in the overall radiation pressure due to the Doppler and Zeeman effects, respectively. MOTs contain from one \cite{singleatom} to $\approx 10^{11}$ \cite{labeyrie, labeyrie2} atoms at milliKelvin typical temperatures, and the atomic species cooled usually has a simple energy level structure -- which covers all alkali metals, alkaline earth species and metastable noble gases. Other atomic species can also be very effectively cooled (e.g. holmium \cite{holmium}), and very recently the first molecular MOT was obtained \cite{molmot}. 

\begin{figure}[!b]
\includegraphics[width=1\columnwidth]{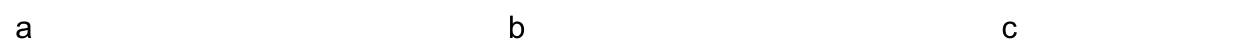}
\centering
\begin{minipage}{.39\columnwidth}
\includegraphics[width=1\columnwidth]{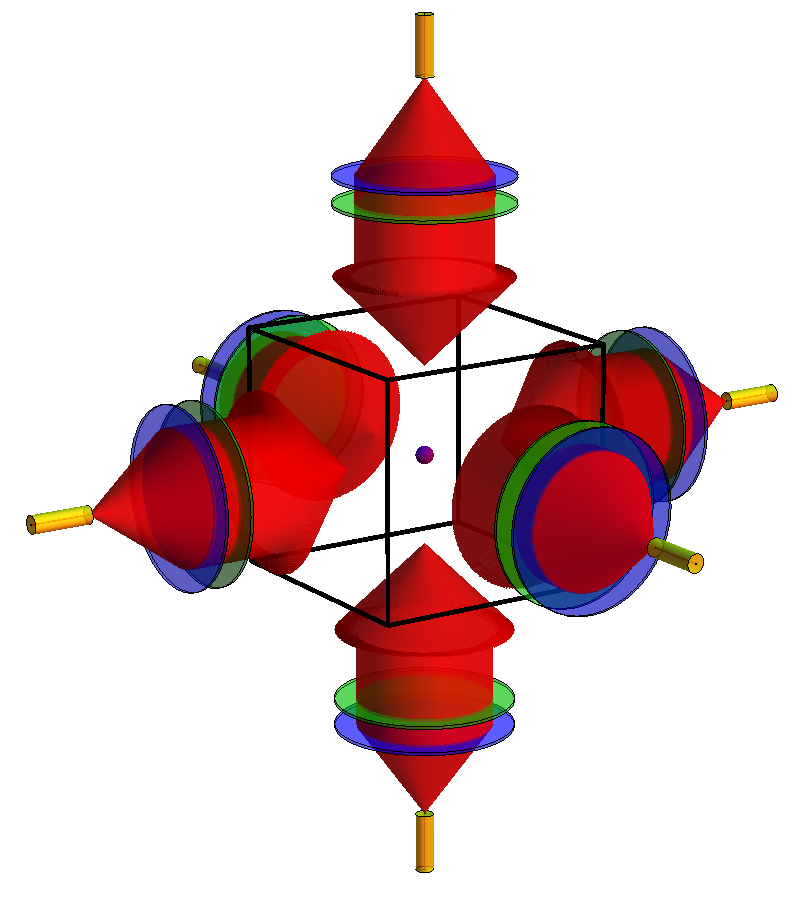}
\end{minipage}
\begin{minipage}{.39\columnwidth}
\includegraphics[width=1\columnwidth]{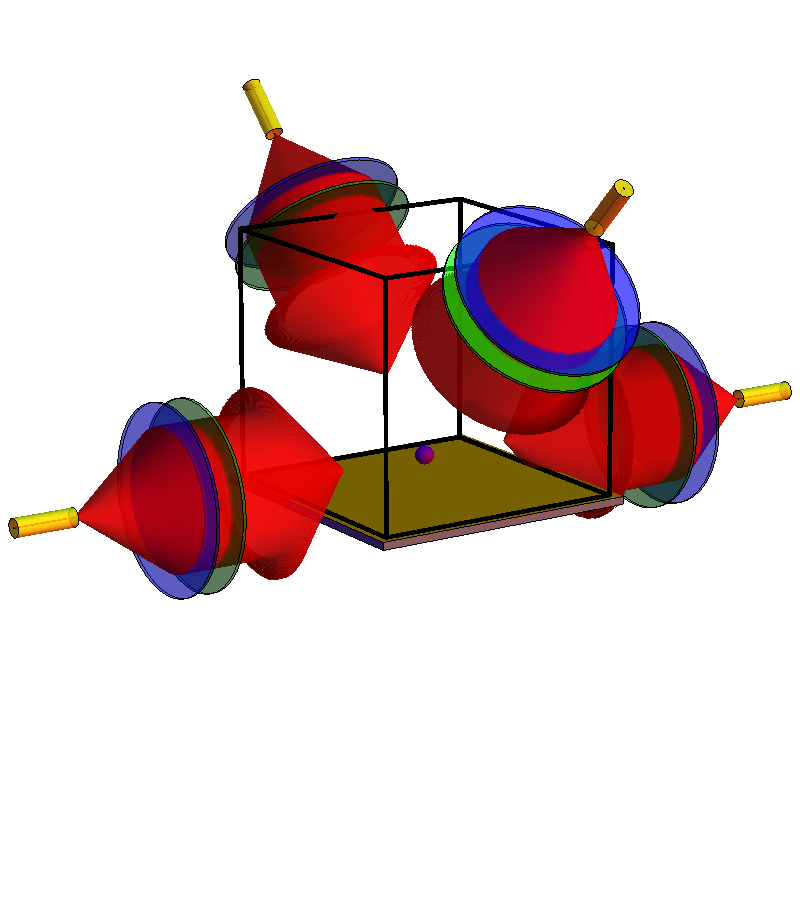}
\end{minipage}
\begin{minipage}{.195\columnwidth}
\includegraphics[width=1\columnwidth]{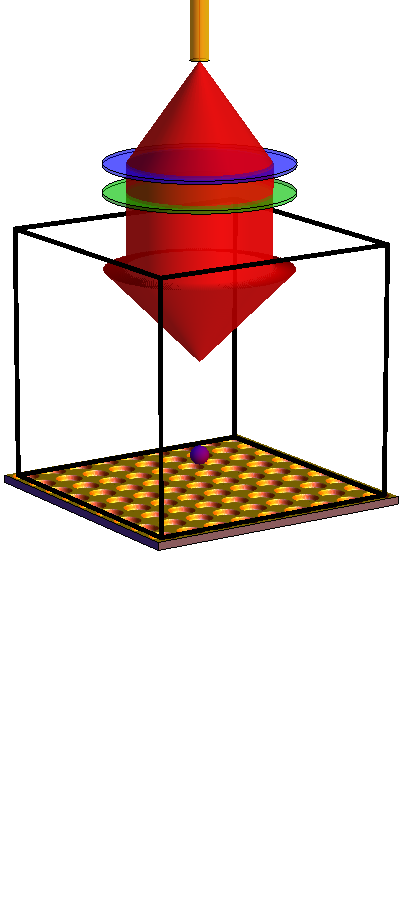}
\end{minipage}
\caption{\label{Fig1}
Magneto-optical trap geometries: six-beam MOT (a), reflection MOT (b) and grating MOT (c). The black cuboid indicates a vacuum cell, yellow fibers deliver cooling and repumping light (red) which is collimated by lenses (blue) and appropriately circularly-polarised with quarter waveplates (green).  The vacuum pump, atom source, anti-Helmholtz magnetic coils etc.\ are omitted for clarity.}
\end{figure}

We have recently highlighted the benefits of using surface-patterned chips to make MOTs with relatively large numbers of ultracold atoms \cite{nshii13}, compared to prior micro-fabricated magneto-optical traps (MOTs) \cite{pollock09,pollock11}. The design requires only the alignment of a single input laser beam, and extends our previous work on tetrahedral mirror \cite{vangeleyn09} and grating \cite{vangeleyn10} geometries, which are four-beam \cite{shimizu} equivalents of pyramidal MOTs \cite{lee96}. Figure~\ref{Fig1} depicts a `like-for-like' graphical comparison of the optical setup in a standard 6-beam MOT (Fig.~\ref{Fig1}(a) \cite{raab87,lind,schwartz}), a mirror MOT (Fig.~\ref{Fig1}(b) \cite{reichel99,hansel01}) and a grating MOT (GMOT, Fig.~\ref{Fig1}(c) \cite{nshii13}). 

\begin{figure}[!t]
\centering
\includegraphics[width=.5\columnwidth]{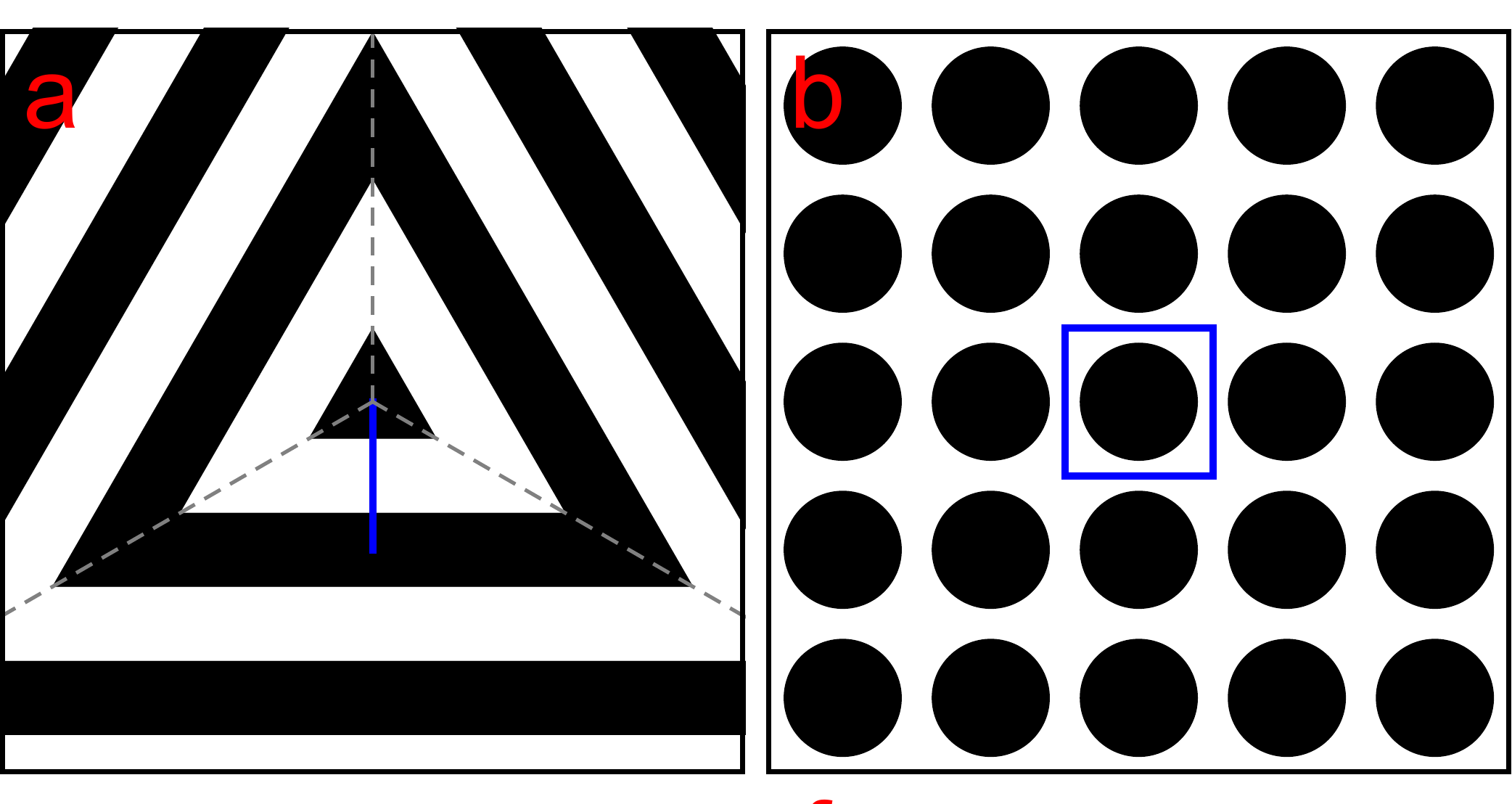}
\caption{\label{Fig2}
The `zoomed in' microstructure on the different $2\times 2\,$cm$^2$ chips used in our studies. The black and white zones are separated by a height difference of $\sim 200\,$nm, one quarter of the wavelength of the rubidium cooling light ($780\,$nm). Chips TRI$_{15}$ and TRI$_{12}$ have the pattern shown in (a), comprising three one-dimensional gratings with period $d$ (blue line) of $1470$ and $1200\,$nm, respectively. The other chip, CIR, has the pattern shown in (b), with a grating unit cell (blue square) side length $d=1080\,$nm. First order Bragg-diffracted beams have angles 32$^{\circ}$ (TRI$_{15}$), 41$^{\circ}$ (TRI$_{12}$), 46$^{\circ}$ (CIR) with respect to the chip surface normal.}
\end{figure}

Here we study the atom-number characteristics of four different chip systems, and more complete details of different fabrication methods for the chips are discussed by Cotter \textit{et al.}~\cite{cotter13}. The two chip geometries we used are shown in Fig.~\ref{Fig2}. One chip style (Fig.~\ref{Fig2}(a)) comprises three one-dimensional gratings arranged in a pattern with 3-fold rotational symmetry (we used chips with periods of $d=1470\,$nm and $d=1200\,$nm). The other chip style (Fig.~\ref{Fig2}(b)) is a two-dimensional grating generated with circular regions centred in square units cells of side length $d=1080\,$nm. Hereafter the chips will be referred to as TRI$_{15}$, TRI$_{12}$ (Fig.~\ref{Fig2}(a)) and CIR  (Fig.~\ref{Fig2}(b)). Although we have used chips TRI$_{15}$ and CIR previously (chips B and C in Nshii \textit{et al.}~\cite{nshii13}), purely for number characterisation, here we present density and temperature data for these chips. Moreover, we also investigate MOTs with improved capture efficiency for the 1D structure (i.e.\ the new chip TRI$_{12}$) and improved radiation balance for the 2D structure (chip CIR$_{\textrm{ND}}$: identical to CIR but with an input beam intensity spatially sculpted with a neutral density filter -- for which the rationale is given in Sec.~\ref{int}).

We begin by introducing simple theoretical models for the MOT temperature (Sec.~\ref{dopptemp}) and atom number (Sec.~\ref{doppnum}), then consider the amelioration of any effects caused by imbalanced radiation pressure in grating MOTs (Sec.~\ref{int}). We then present the experimental data (Sec.~\ref{exp}), comparing and contrasting with the theoretical predictions. Finally we present `best of both worlds' findings (Sec.~\ref{bbw}) -- by permitting analog temporal evolution of both MOT intensity and detuning we go beyond Nshii \textit{et al.}~\cite{nshii13} to achieve high atom numbers and sub-Doppler temperatures \textit{simultaneously} using both 1D and 2D grating structures. 

\section{Theory: Doppler temperature}
\label{dopptemp}

Whilst sub-Doppler cold atom temperatures can be reached in optical molasses (Sec.~\ref{bbw}) \cite{subdopp,subdopp2,subdopp3}, the conditions for atom collection in a MOT are in stark contrast to those required to reach the lowest temperatures: a MOT needs high beam intensities relative to the saturation intensity $(I\gg I_S)$, detuning of a few natural linewidths $\Delta\approx -2\Gamma$ and a magnetic gradient $\approx 10\,$G/cm for a typical alkali metal vapour cell MOT with $25\,$mm beam diameter. For this reason the considerably simpler Doppler theory is a good description for our MOT temperature. The standard 3D Doppler result for a sample of atoms uniformly and isotopropically illuminated in three dimensions by light of total intensity $I_{T},$ with corresponding saturation parameter $\beta_T=I_T/I_S,$ and detuning $\Delta$ \cite{optmol} is:
\begin{equation}
T_U = T_D \frac{1+\beta_T+4 \Delta^2/\Gamma^2}{-4 \Delta/\Gamma},\label{doptemp}
\end{equation}
where $T_D=\frac{\hbar \Gamma}{2 k_B}.$
In our $^{87}$Rb experiment we cool using the rubidium D2 transition which has a natural linewidth $\Gamma= 2 \pi \times 6.07\,$MHz. The standard Doppler temperature minimum in the limit of low intensity and detuning $\Delta=-\Gamma/2$ is $T_D$, i.e. $146\,\mu$K for rubidium, with the more general lowest temperature $T_D \sqrt{1+\beta_T}$ minimised at a detuning $\Delta=-\sqrt{1+\beta_T}\;\Gamma/2$. Throughout this paper, for consistency, we use the saturation intensity averaged over all polarisations and magnetic sub-levels $I_S=3.57\,$mW/cm$^2$ -- however we note that using the stretched state saturation intensity $I_S=1.67\,$mW/cm$^2$ for the theoretical temperature of Eq.~(\ref{doptemp}) leads to much better agreement with our experimentally measured temperatures (Sec.~\ref{exp}).

Although the velocity distribution in a magneto-optical trap is Gaussian, the system is by no means in thermal equilibrium in the usual sense, with the typical light scattering rate from atoms vastly larger than the interatomic elastic collision rate. The final `temperature' in a given direction is given by the balance between the heating (cf. diffusion coefficient and light scattering rate) and the velocity damping constant $\gamma$. In our system the effective temperature parallel to and perpendicular to the grating are therefore expected to be different.

For our gratings if the incident beam has a spatially uniform intensity $I_0$ and wavevector $\textbf{k}_0=k\{0,0,-1\}$, for balanced optical pressure from $N$ diffracted orders each should ideally carry an intensity of $I_i=I_0/(N \cos\alpha)$ with $i\in \{1,...,N\}$. The wavevectors of the diffracted beams are $\textbf{k}_i=k\{\sin\alpha\sin\frac{2 \pi i}{N},\sin\alpha\cos\frac{2 \pi i}{N},\cos\alpha\}$, where $\alpha$ is the angle the diffracted orders make to the grating normal $\{0,0,1\}$ ($\alpha=2\theta$ cf.\ \cite{vangeleyn09,vangeleyn10}). The total intensity is therefore $I_T=\sum_{i=0}^{N}{I_i}=(1+\sec\alpha)I_0,$ with each beam of intensity $I_i$ contributing a relative intensity-dependent heating proportional to its intensity along the beam direction due to absorption, and an equal associated heating rate due to spontaneous emission. The heating from spontaneous emission is assumed for simplicity to be isotropic, with the heating rate for light with a wavevector $\textbf{k}_i=\{k_{i_x},k_{i_y},k_{i_z}\}$ proportional to $\hbar^2 R_i \,k^2\{\frac{1}{3},\frac{1}{3},\frac{1}{3}\}$ in the $x, y$ and $z$ dimensions, for a scattering rate
\begin{equation}
R_i(\textbf{v})=\frac{\Gamma}{2} \frac{I_i/I_S}{1+\beta_T+4(\Delta-\textbf{k}_i \cdot \textbf{v})^2 /\Gamma^2}, \textrm{with }R_i(\textbf{0})=\frac{I_i}{I_S} R=\beta_i R.
\end{equation}
Assuming dipolar (as opposed to isotropic) spontaneous emission from a circularly polarised atom adds complexity and does not significantly alter the predicted temperatures. 
The heating due to absorption is apportioned $\hbar^2 R_i \{{k_{i_x}}^2,{k_{i_y}}^2,{k_{i_z}}^2\}$ in the three spatial dimensions. The total heating rate near zero velocity in each dimension is thus:
\begin{equation}
\!\!\!\!\!\!\!\!\!\!\!\!\!\!\!\!\!\!\!\!\!\!\!\!\!\textbf{D}=\sum_{i=0}^{N}{\hbar^2 \beta_i R \left\{{k_{i_x}}^2+\frac{k^2}{3},{k_{i_y}}^2+\frac{k^2}{3},{k_{i_z}}^2+\frac{k^2}{3}\right\}}
\end{equation}
which simplifies to 
\begin{equation}
\!\!\!\!\!\!\!\!\!\!\!\!\!\!\!\!\!\!\!\!\!\!\!\!\! \textbf{D}=\hbar^2 k^2 \frac{I_0}{6 I_S} R \left( \left\{ 2,2,8\right\} +\sec \alpha \left\{3 \sin^2\alpha+2, 3 \sin^2\alpha+2, 6\cos^2\alpha+2 \right\}\right)\label{diff}
\end{equation}

In balanced laser cooling the relative damping constants can be determined by the Taylor expansion of the total force equation $\textbf{F}=\sum_{i=0}^{N}{\hbar \textbf{k}_i R_i(\textbf{v})}$ about $\textbf{v}=\textbf{0}.$ The damping constants in the $xy$ directions ($\gamma_\parallel$ parallel to the grating) and $z$ direction ($\gamma_\perp$ perpendicular to the grating) are thus given by \cite{vangeleyn09}:
\begin{eqnarray}
\!\!\!\!\!\!\!\!\!\!\!\!\!\!\!\!\textbf{F}(\textbf{v}) & \approx & - \boldmath{\gamma}\cdot \textbf{v}=-\{\gamma_\parallel v_x,\gamma_\parallel v_y,\gamma_\perp v_z\}\nonumber \\
\!\!\!\!\!\!\!\!\!\!\!\!\!\!\!\! \{\gamma_\parallel,\gamma_\perp\} &=&\gamma_6 \left\{\frac{\sin\alpha\tan\alpha}{4},\frac{1+\cos\alpha}{2}\right\},\;\;\textrm{where }\gamma_6=\frac{-16 \hbar I_0 k^2 R \Delta}{I_S \Gamma^2 (1+\beta_T+\frac{4\Delta^2}{\Gamma^2})}\label{damp}
\end{eqnarray}
 is the damping force in a standard 6-beam MOT.
The temperature in each dimension can then be determined as proportional to the ratio of the diffusion \textbf{D} (Eq.~(\ref{diff})) to the damping $\mathbf{\gamma}$ (Eq.~(\ref{damp})):
\begin{equation}
\!\!\!\!\!\!\!\!\!\!\!\!\!\!\!\!\{T_\parallel,T_\perp\}=\frac{T_U}{6}\left\{ 3+\csc^2(\alpha/2), 3+\sec\alpha \right\},
\end{equation}
which reduces to the standard isotropic temperature $T_U$ when $\alpha=\arccos(1/3).$ Note this temperature only depends on the grating diffraction angle and not on the number of diffracted orders $N.$

The temperatures obtained experimentally (Sec.~\ref{exp}) do not reflect the theoretical temperature disparity in the directions parallel and perpendicular to the grating. The model above only includes forces due to absorption of light \textit{directly} from the laser beams, however for MOTs with larger atom number and light scattering rate (like in our experiment), there will be a significant contribution to the forces in the MOT due to the $1/r^2$ repulsion from closely packed atoms with inter-particle spacing $r$ due to absorption of spontaneously emitted light from neighbouring atoms \cite{rerad,rerad2,rerad3,rerad4}. These forces effectively mix the energies across dimensions, creating a more uniform temperature distribution. The theoretical temperature we use in both dimensions is therefore $T=\frac{2}{3}T_\parallel+\frac{1}{3}T_\perp$, which agrees surprisingly well with experimental values given the simplicity of the model.     

For completeness and to determine theoretical MOT density, one can also derive the spring constant $\kappa$ of the trap, expanding the total force $\textbf{F}$, but this time relative to position. The $\alpha$ dependencies of the damping constants are 
$\{\kappa_\parallel,\kappa_\perp\}\propto\{\frac{1}{4}\sin\alpha\tan\alpha,1-\cos\alpha\},$ the same as derived by Vangeleyn \textit{et al.} in \cite{vangeleyn09}. The damping therefore reduces to an isotropic $\frac{2}{3}$ ($\frac{1}{3}$) of the standard 6-beam MOT radial (axial) $\kappa$ values, if the GMOT has $\alpha=\arccos(1/3)$ and laser beams of the same intensity as the standard MOT. Under these conditions the GMOT should be spatially isotropic -- it will have equal rms cloud radii as a standard MOT in both directions parallel to the grating, but twice the standard MOT extent in the direction perpendicular to the grating. An important caution against using this Doppler spring constant to determine density, however, is that the experimental spring constant of a trap is notoriously difficult to determine, and the reradiative forces \cite{rerad,rerad2,rerad3,rerad4} discussed in the previous paragraph make simple models of the effective spring constant considerably more difficult. 

\section{Theory: Doppler atom number}
\label{doppnum}

In order to predict the number of atoms in the MOT, we use the standard method of Lindquist \textit{et al.} \cite{lind}: treat the MOT as a spherical target region with radius $r$ and cross-section $4 \pi r^2$, then find the flux of atoms incident on the target, with speeds less than the capture velocity $v_c$ of the MOT slowing region. In the ideal case of a rubidium-dominated vapour both loading and loss mechanisms are proportional to atomic density, which is therefore irrelevant for calculating atom number. It is necessary to know the collisional cross section of cold atoms as seen by hot atoms, which for rubidium is $\sigma=2 \times 10^{-17}\,$m$^2$ \cite{gould} (similar to caesium \cite{lind}). We make the assumption of a 1D optical molasses from two counter-propagating laser beams. The total acceleration is modified by a prefactor $\eta$ to allow for geometric effects:
\begin{eqnarray}
a=v \frac{dv}{dx}&=& \eta \frac{h k \Gamma I}{2 m I_S}\left( \frac{1}{1+\frac{I_T}{I_S}+4 \frac{(\Delta -k v)^2}{\Gamma^2}}- \frac{1}{1+\frac{I_T}{I_S}+4 \frac{(\Delta +k v)^2}{\Gamma^2}}\right).
\end{eqnarray}
By rearranging one arrives at an equation of the form $f(v) dv= B dx$, where $f(v)$ is a quartic polynomial in $v$ and $B$ is a constant. This equation can then be integrated to analytically determine $x(v)$. The integration constant is chosen such that $x(0)=0$. Setting $x(v_c)=2 r$ (atoms are stopped over a distance of twice the MOT radius) and subsequently inverting gives $v_c$, the capture velocity. The total steady-state number of atoms in the MOT is then \cite{lind}:
\begin{equation}
N=\frac{4 \pi r^2}{8 \sigma} \left( \frac{v_c}{v_T} \right)^4,
\end{equation}
where $v_T=\sqrt{2 k_B T/m}$ is a thermal velocity in the (hot) background vapour.

We note that the saturation intensity features in the theoretical temperature and both theoretical and experimental atom number, and we therefore consistently use the saturation intensity averaged over all polarisations and magnetic sub-levels $I_S=3.57\,$mW/cm$^2$, which is valid for small detunings \cite{aaphd,booth}. There are, however, arguments for using a value closer to the stretched state saturation intensity $I_S=1.67\,$mW/cm$^2$ \cite{footpsd}. We used this value in our previous work on atom number \cite{nshii13} as a precaution to prevent over-counting the atom number, although for high levels of saturation the effect of changing $I_S$ on experimental atom number is negligible.  

\section{Theory: Non-uniform beam intensity}
\label{int}

In the theory so far we have assumed a spatially uniform incident laser intensity. However, real propagating laser beams tend to have Gaussian spatial intensity profiles. If such intensity distributions are apertured, or their intensity is spatially modified in other respects, as long as the modifications occur on large (centimetre size) beams, diffraction effects can be neglected. While various refractive beam shapers which transform Gaussian beams into flat-top intensity beams are available on the market, these tend to be expensive (albeit power efficient) and designed for relatively small output beam diameters. Spatially shaping a Gaussian beam with radial profile $I=I_0 \exp(-2 r^2/w^2)$ using an apodizing filter of outer radius $r_a$ with a neutral density $\textrm{ND}=\exp(2 (r^2-{r_a}^2)/w^2)$ leads to a uniform beam intensity, with reduced intensity $I_0 \exp(-2 {r_a}^2/w^2)$ across the beam. The relative power in the beam is then $2 (r_a/w)^2 \exp(-2 {r_a}^2/w^2)$ which has a maximum of $1/e \approx 37\%$ when the aperture radius is $r_a=w/\sqrt{2}$ -- i.e. even under optimal conditions it is a fairly lossy solution.

\begin{figure}[!b]
\centering
\begin{minipage}{.48\columnwidth}
\includegraphics[width=\columnwidth]{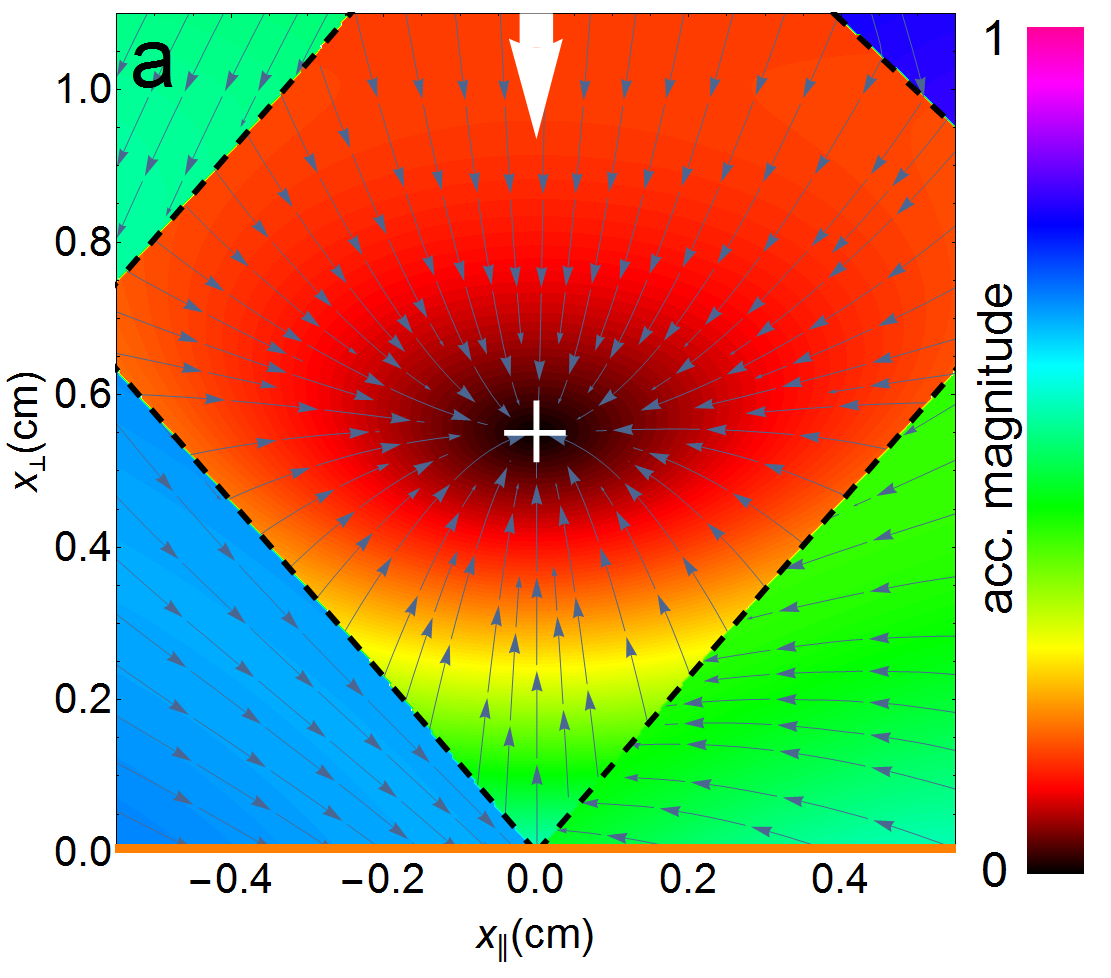}
\end{minipage}
\begin{minipage}{.48\columnwidth}
\includegraphics[width=\columnwidth]{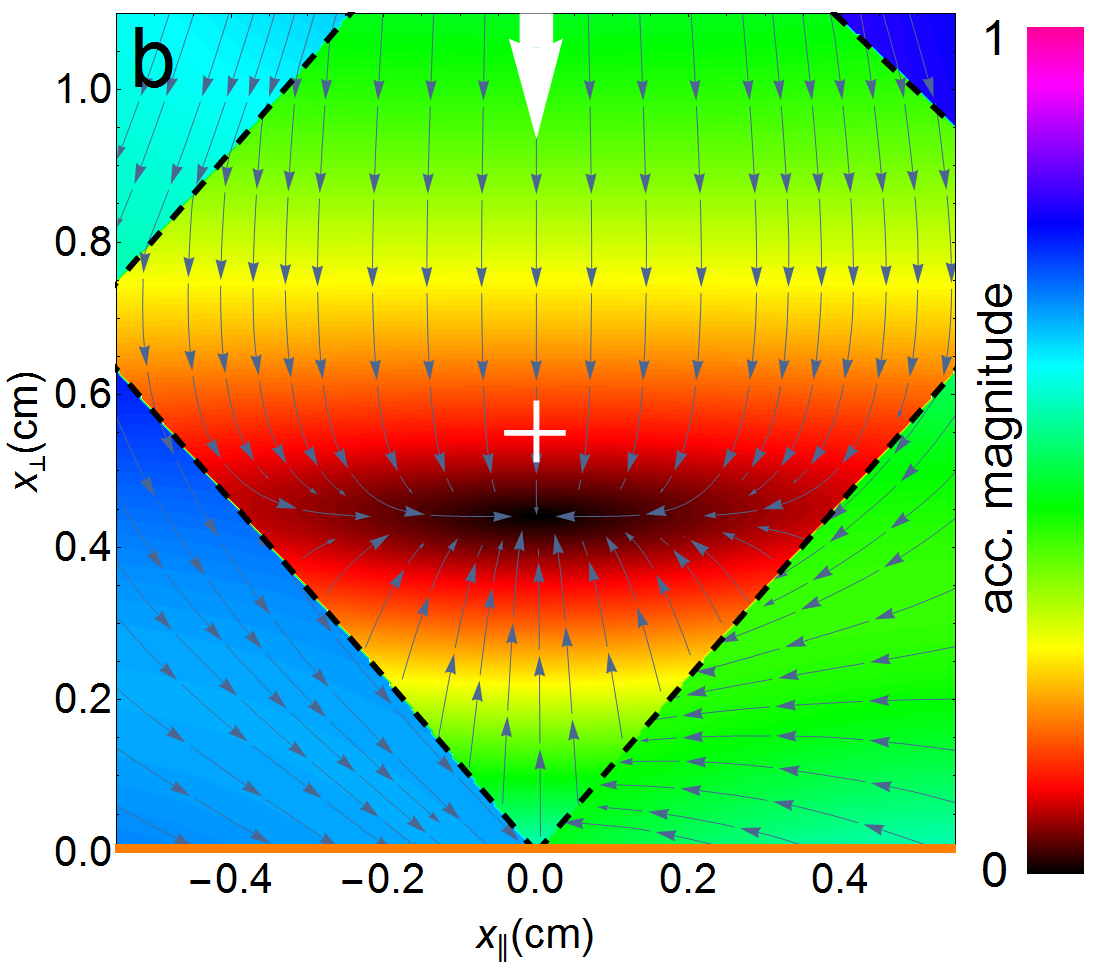}
\end{minipage}
\caption{The full theoretical acceleration above chip TRI$_{12}$, shown both as vector streamlines and magnitude. The white arrow indicates the centre and direction of the beam incident on the grating (gold line at figure base), whereas the white cross indicates the quadrupole magnetic field centre. In figure a, a uniform beam intensity profile is used, whereas in b we use a Gaussian beam with $1/e^2$ radius of $2\,$cm, like in the experiment. Black dashed lines indicate the boundaries of the beam overlap (capture) volume.\label{FigImbal}}
\end{figure}

To determine the effect of Gaussian intensity profiles, we fully integrate the acceleration from a MOT with a single input beam and 3 diffracted orders at diffraction angle $41^\circ$, just like Chip TRI$_{12}$ (Fig.~\ref{FigImbal}). We assume similar parameters to the MOT experiment -- a Gaussian input trap beam with intensity $10\,$mW/cm$^2$ and detuning $\Delta/\Gamma=-1.7$ apertured to a $12\,$mm radius and a magnetic quadrupole with axial gradient $15\,$G/cm. In situations where the beam intensity is spatially uniform (Fig.~\ref{FigImbal}(a)) the radiation pressure balance is uniform throughout the beam-overlap volume. However, with a Gaussian beam intensity profile (Fig.~\ref{FigImbal}(b)), the diffracted orders at the MOT location originate from lower and lower intensity regions of the Gaussian beam as the MOT position is raised from the grating surface. This means that the MOT is pushed downwards more, relative to the magnetic quadrupole centre, as the MOT is raised above the grating surface. Moreover, there is a marked change in the relative trapping and damping constants: $\{\kappa_\parallel,\kappa_\perp,\gamma_\parallel,\gamma_\perp,\}=\{0.21,0.37,0.21,1.31\}$ in Fig.~\ref{FigImbal}(a) and $\{0.10,0.49,0.16,0.96\}$ in Fig.~\ref{FigImbal}(b)  (compared to the isotropic values of $\{1,1,1,1\}$ for $\alpha=\textrm{arccos}(1/3)$ and uniform intensity).

The case of uniform intensity is preferable, as long as one has optimal diffraction efficiency in the gratings, as then the MOT and molasses properties do not depend on MOT position. However, if one uses Gaussian beams and the diffraction efficiency is too high, one can compensate by raising the MOT position to a location where optical forces are balanced and suitable for sub-Doppler cooling. In Sections~\ref{exp} and \ref{bbw} with chip CIR$_\textrm{ND}$ we show how locally shaping the input beam intensity can dramatically improve both MOT and molasses performance.

\section{Experiment}  
\label{exp}

The experimental conditions were very similar to those detailed in the Methods section of Nshii \textit{et al.} \cite{nshii13}.  One major difference is that we now have analogue as opposed to binary intensity control, using an acousto-optic modulator (AOM) after our tapered amplifier prior to fiber-coupling. After the MOT reached steady-state atom number, the atoms were ballistically released for $0,$ $1.5$ or $2\,$ms, and Gaussian fits were used to extrapolate cloud rms radii and thereby MOT temperatures. To build up statistics each drop time was iterated five times. For intensities above and below $10\,$mW/cm$^2$ the fluorescent imaging exposure times were $430\,\mu$s and $150\,\mu$s, respectively.


\begin{figure}[!p]
\centering
\begin{minipage}{.3835\columnwidth}
\includegraphics[width=\columnwidth]{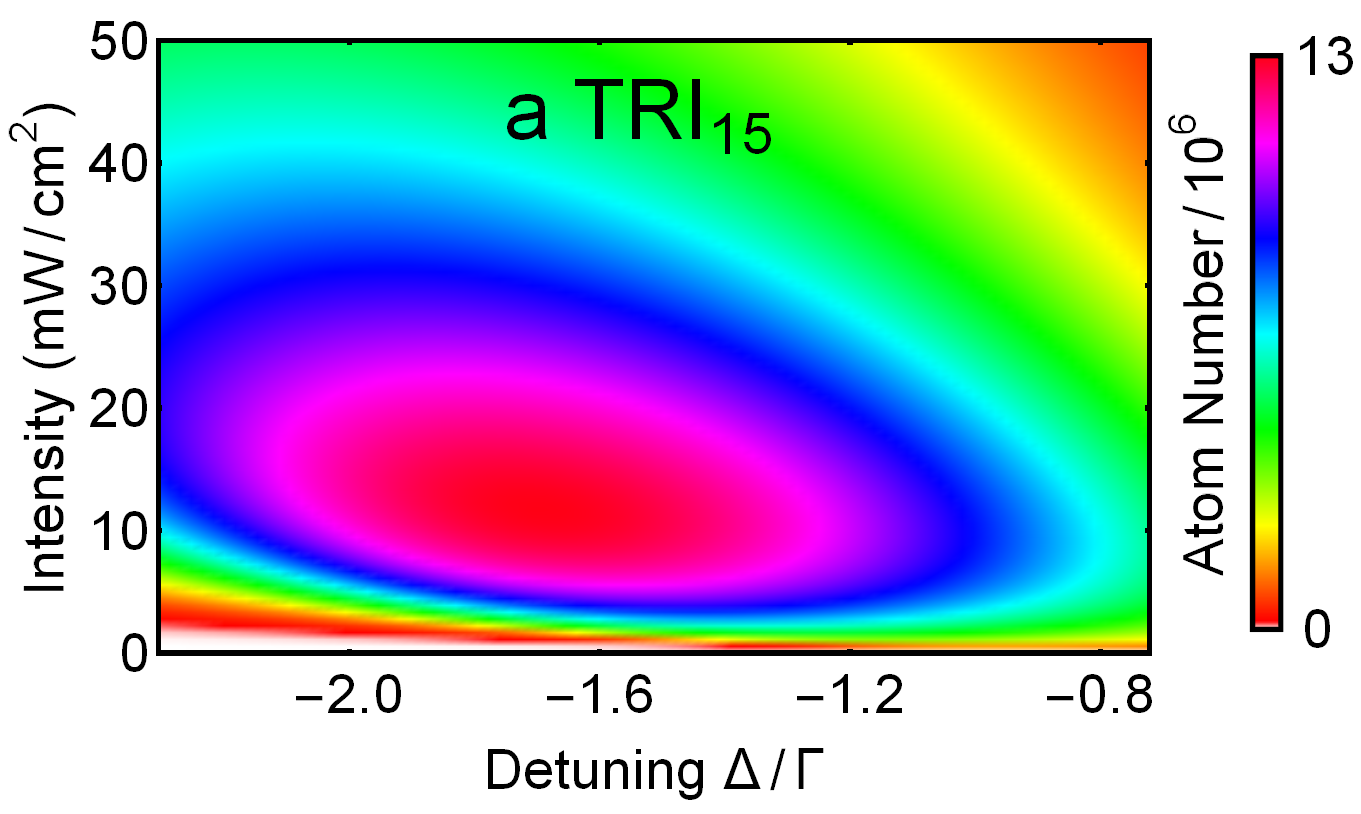}
\includegraphics[width=\columnwidth]{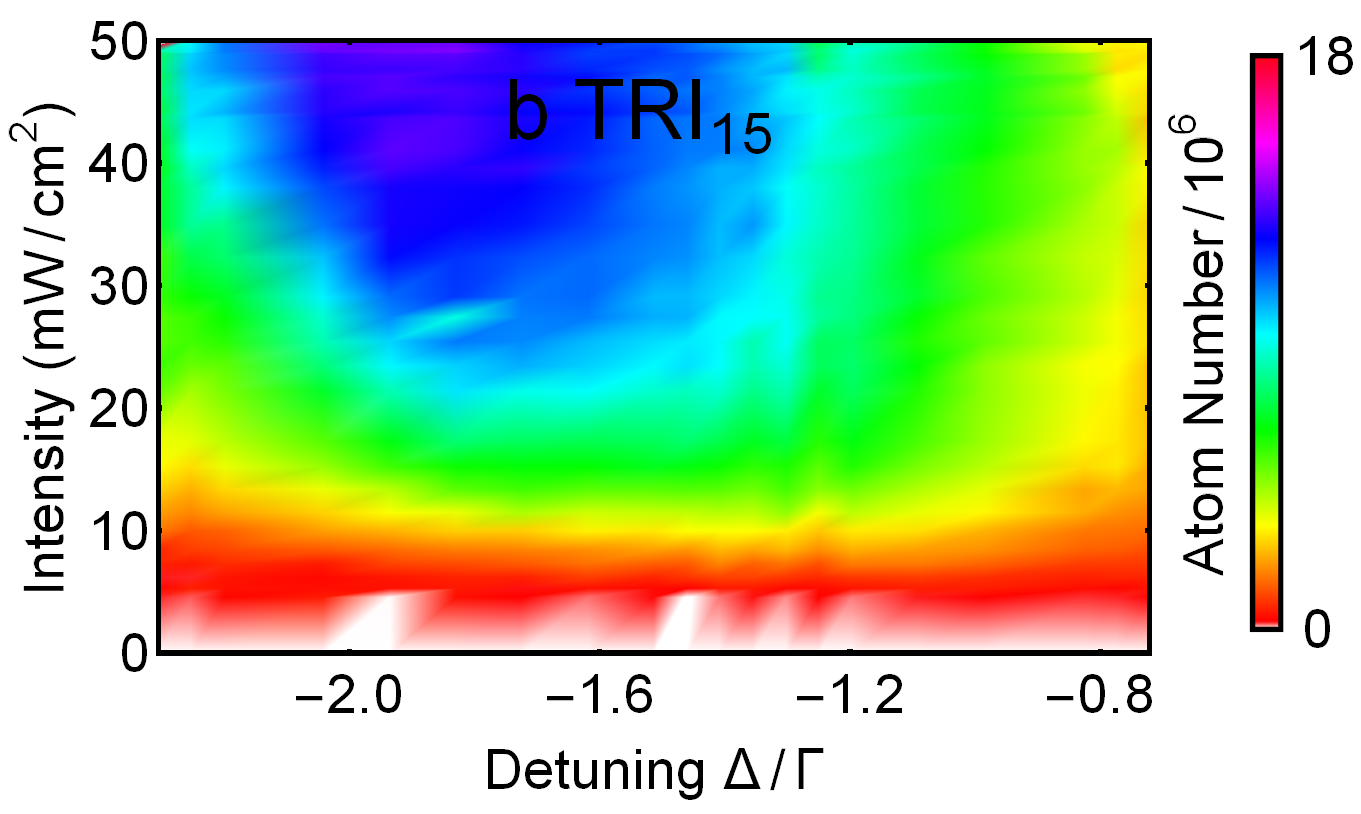}
\includegraphics[width=\columnwidth]{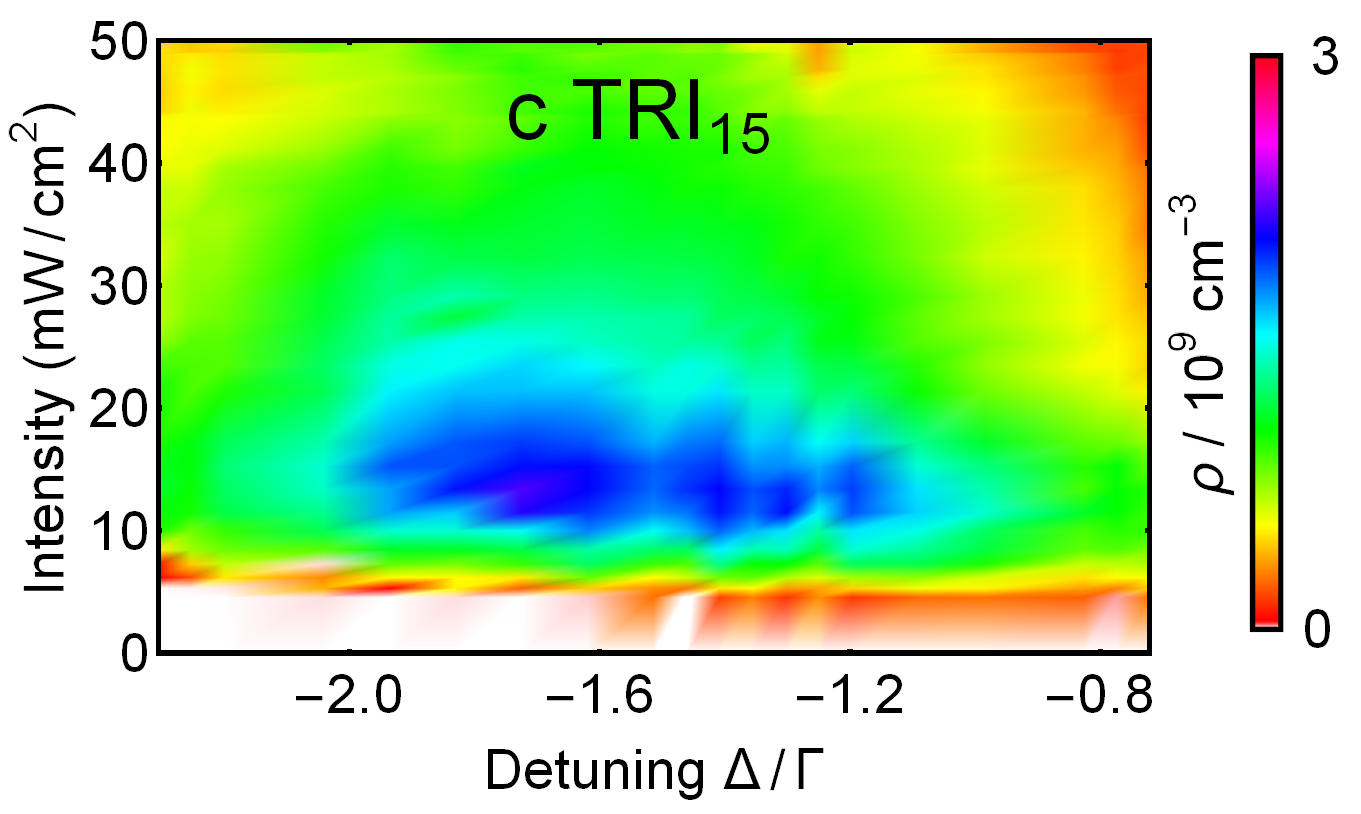}
\includegraphics[width=\columnwidth]{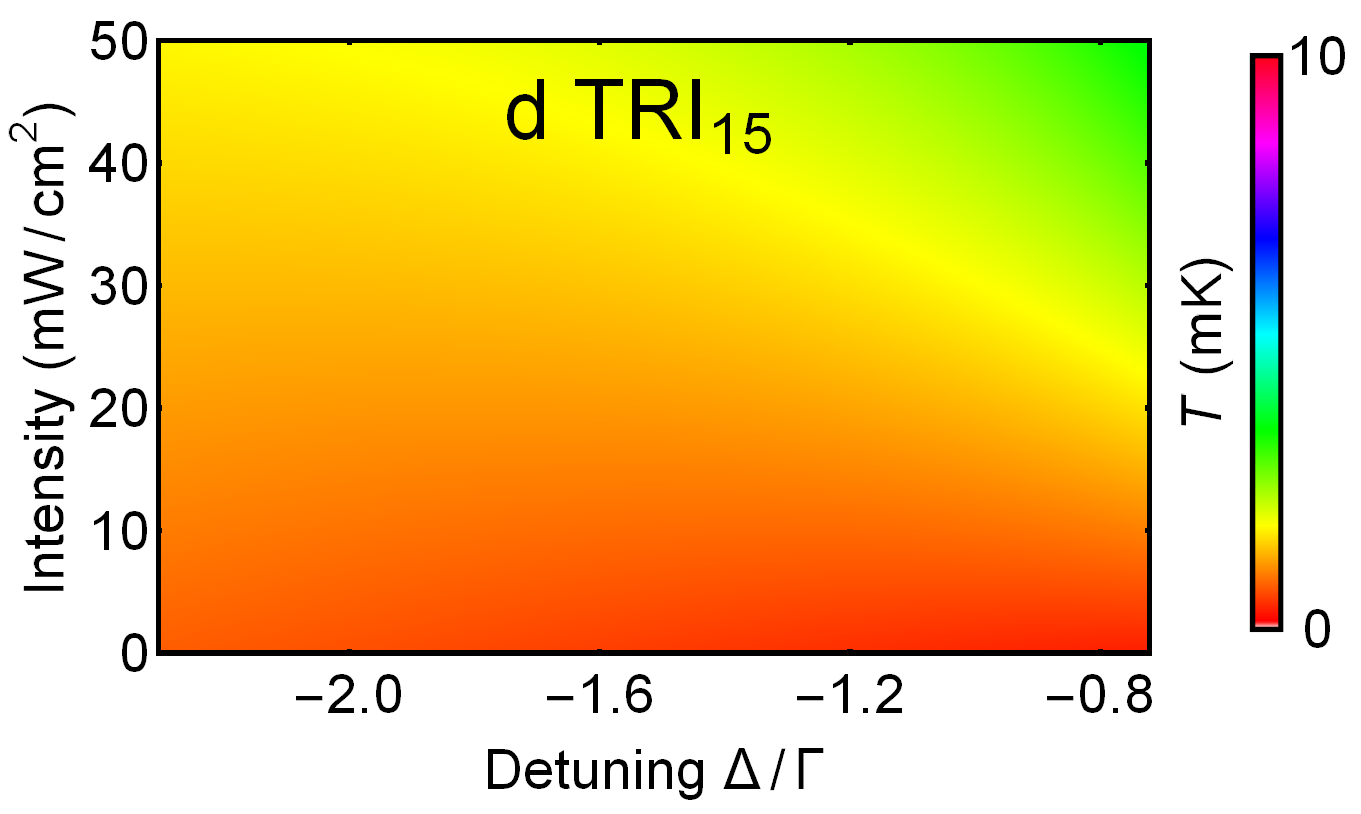}
\includegraphics[width=\columnwidth]{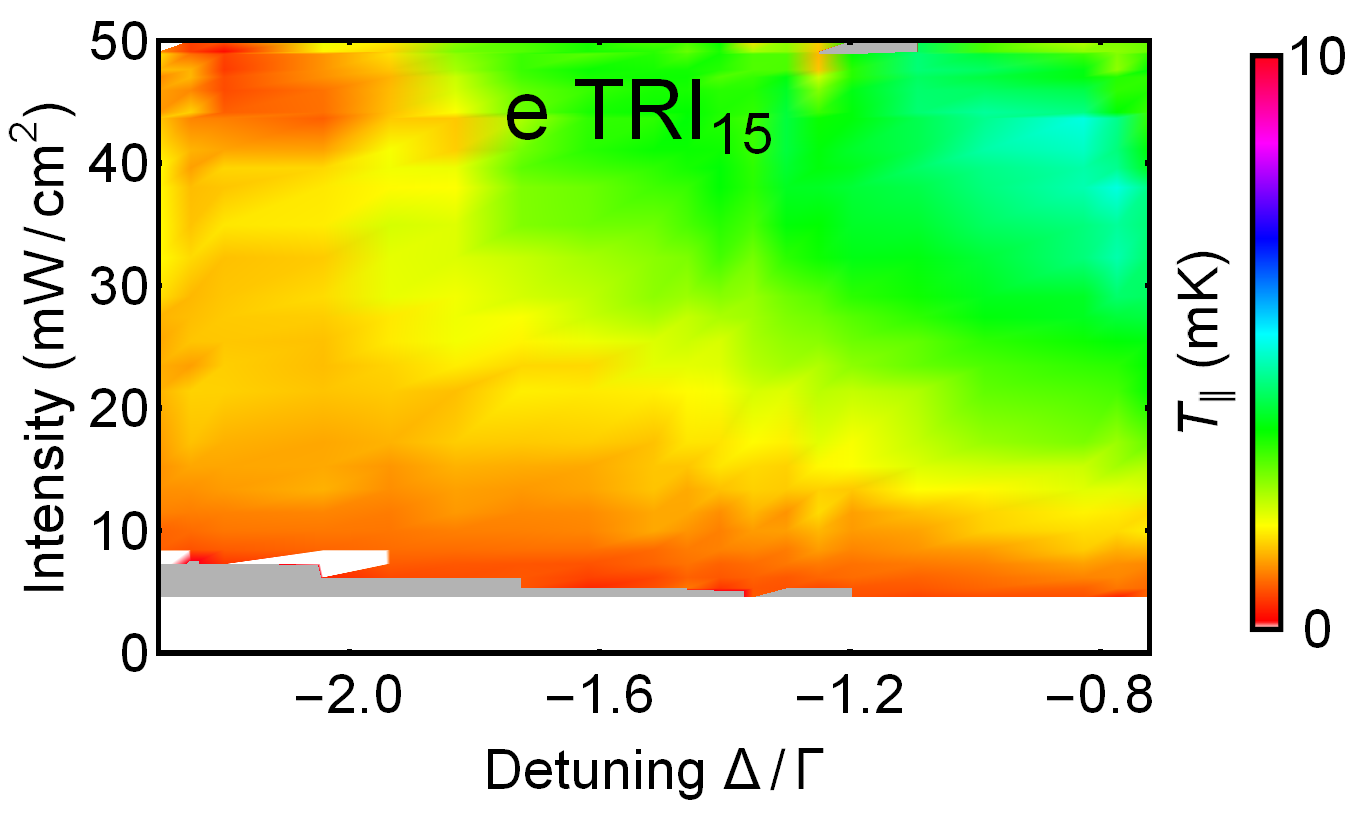}
\includegraphics[width=\columnwidth]{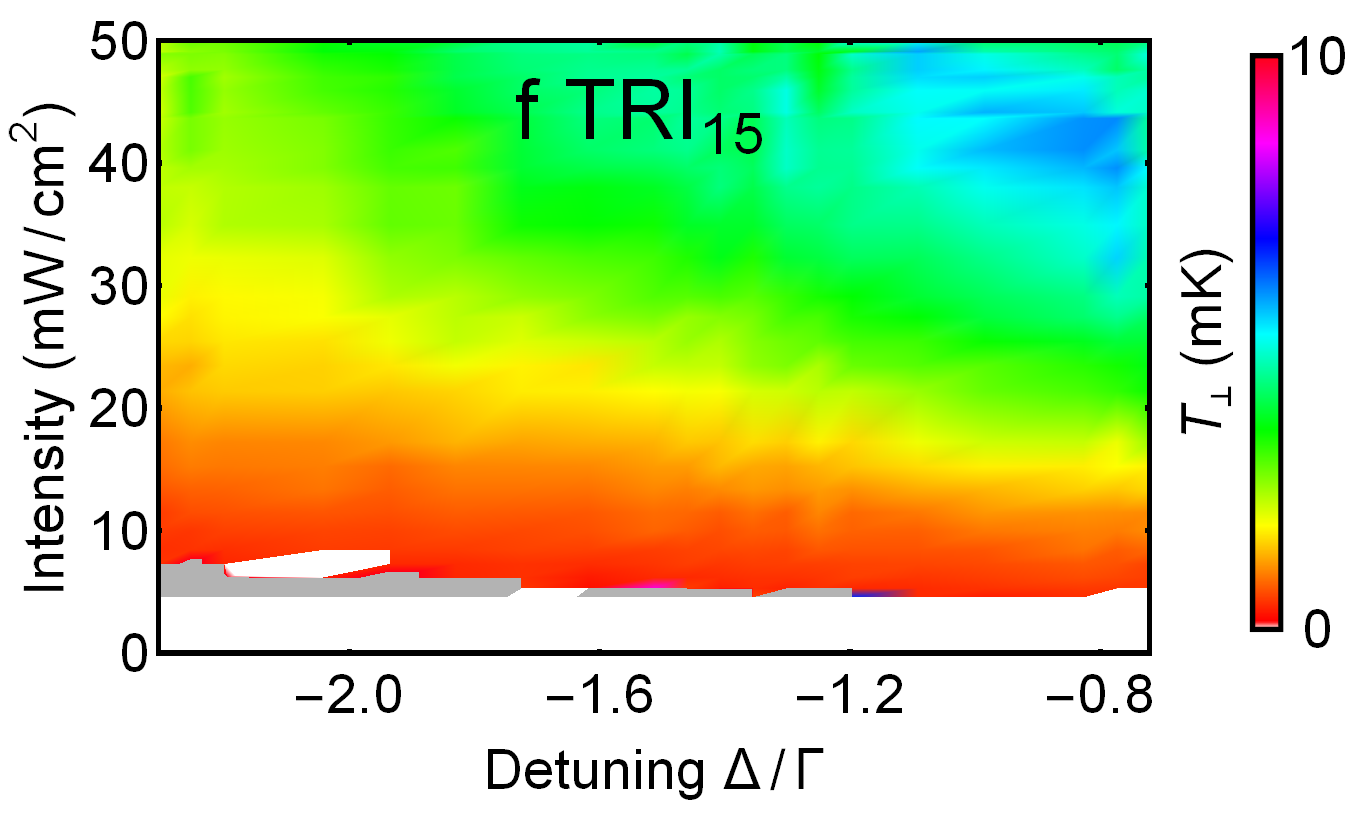}
\end{minipage}
\begin{minipage}{.3835\columnwidth}
\includegraphics[width=\columnwidth]{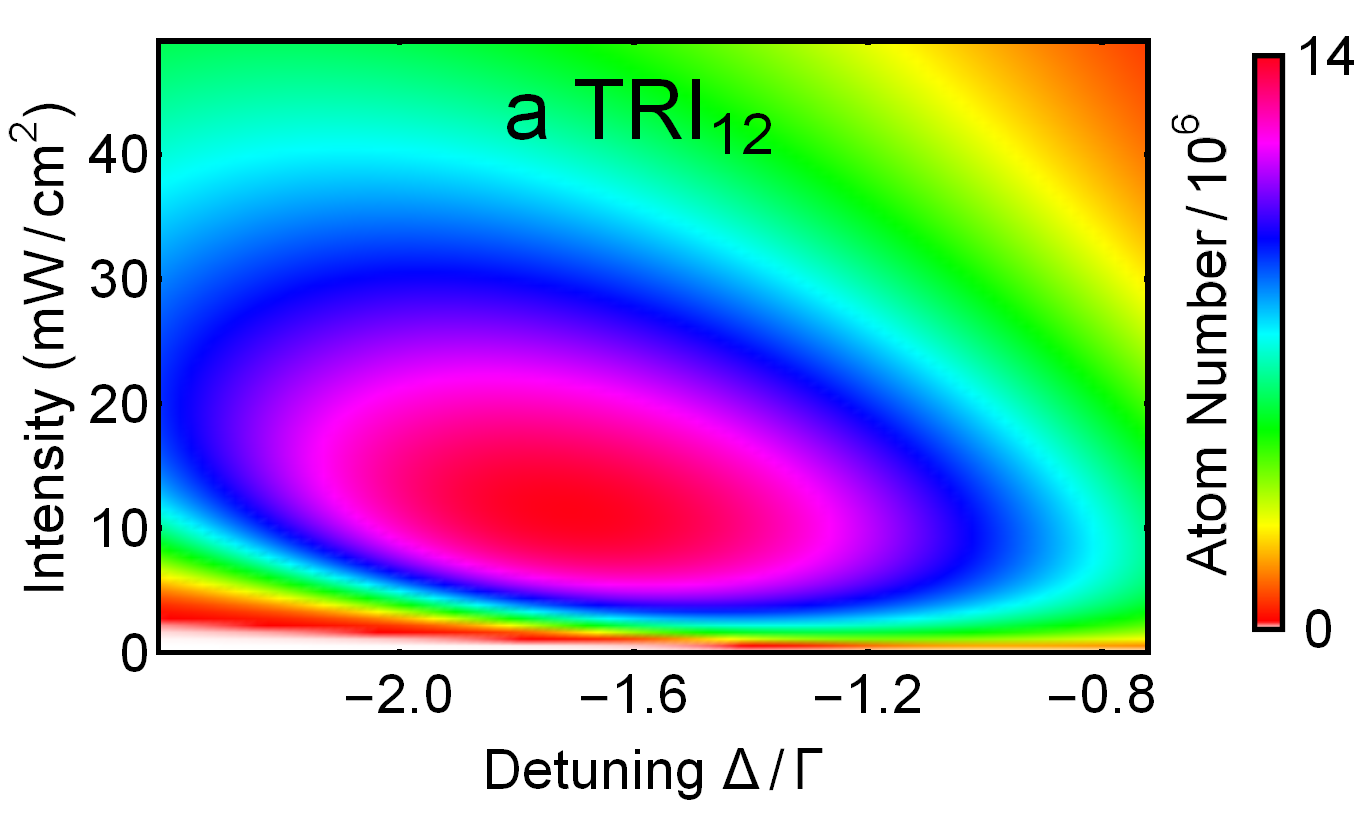}
\includegraphics[width=\columnwidth]{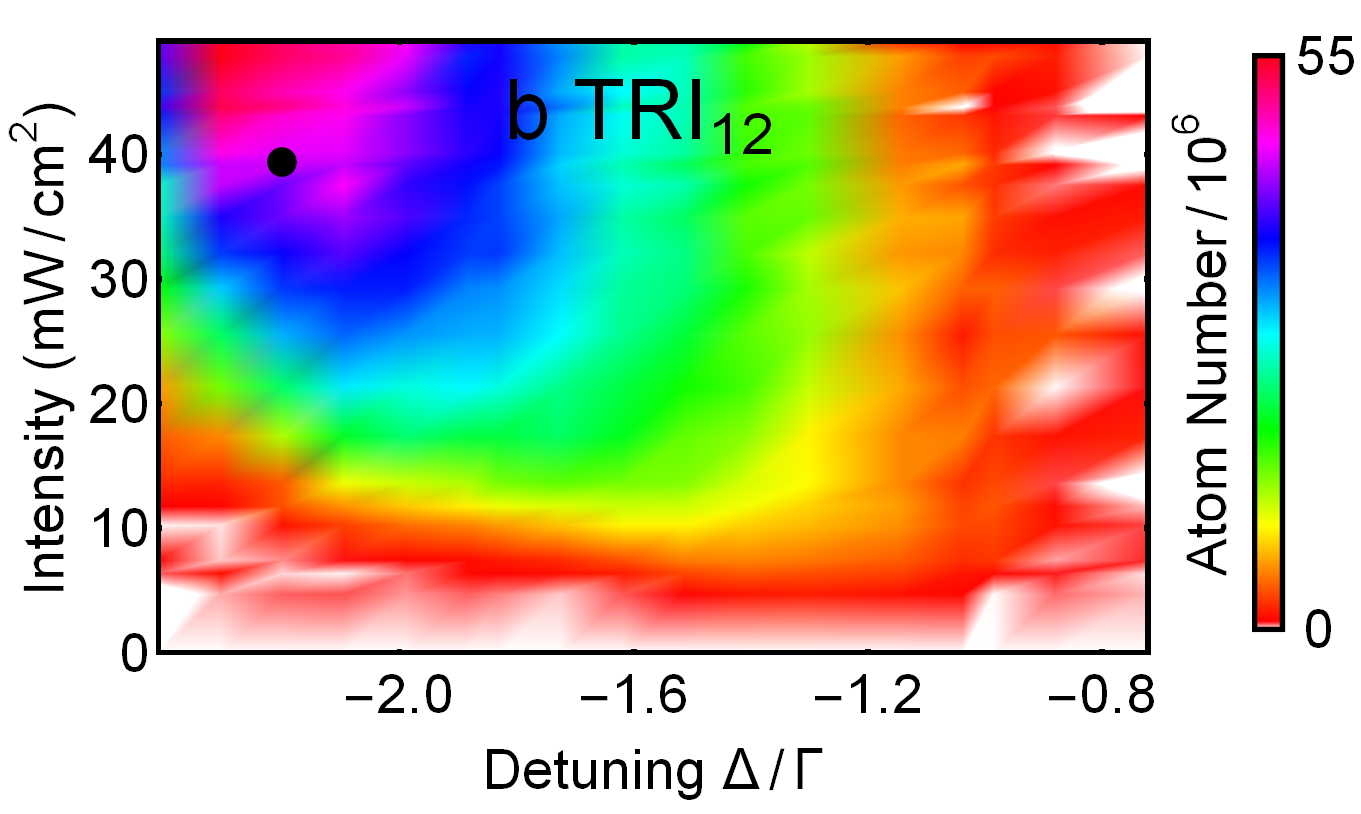}
\includegraphics[width=\columnwidth]{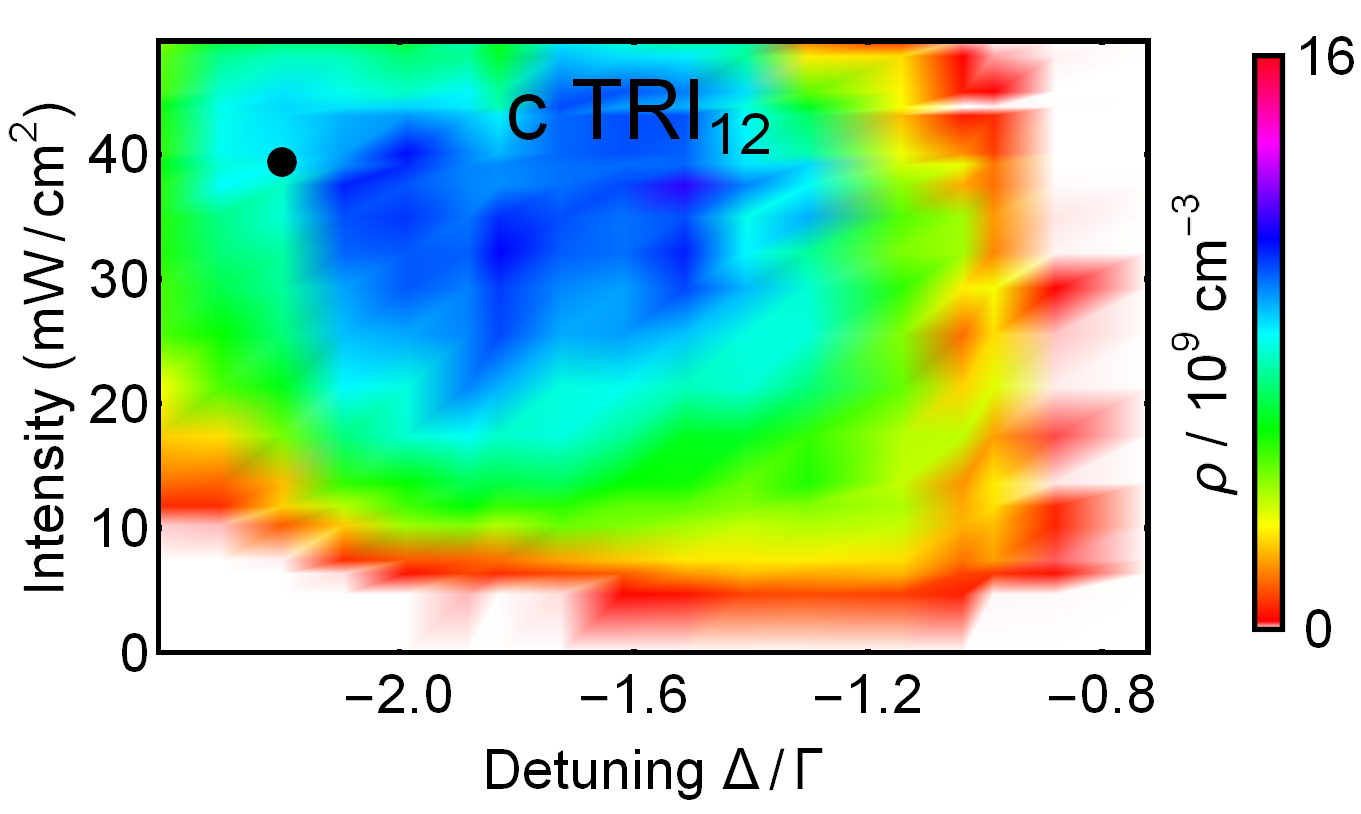}
\includegraphics[width=\columnwidth]{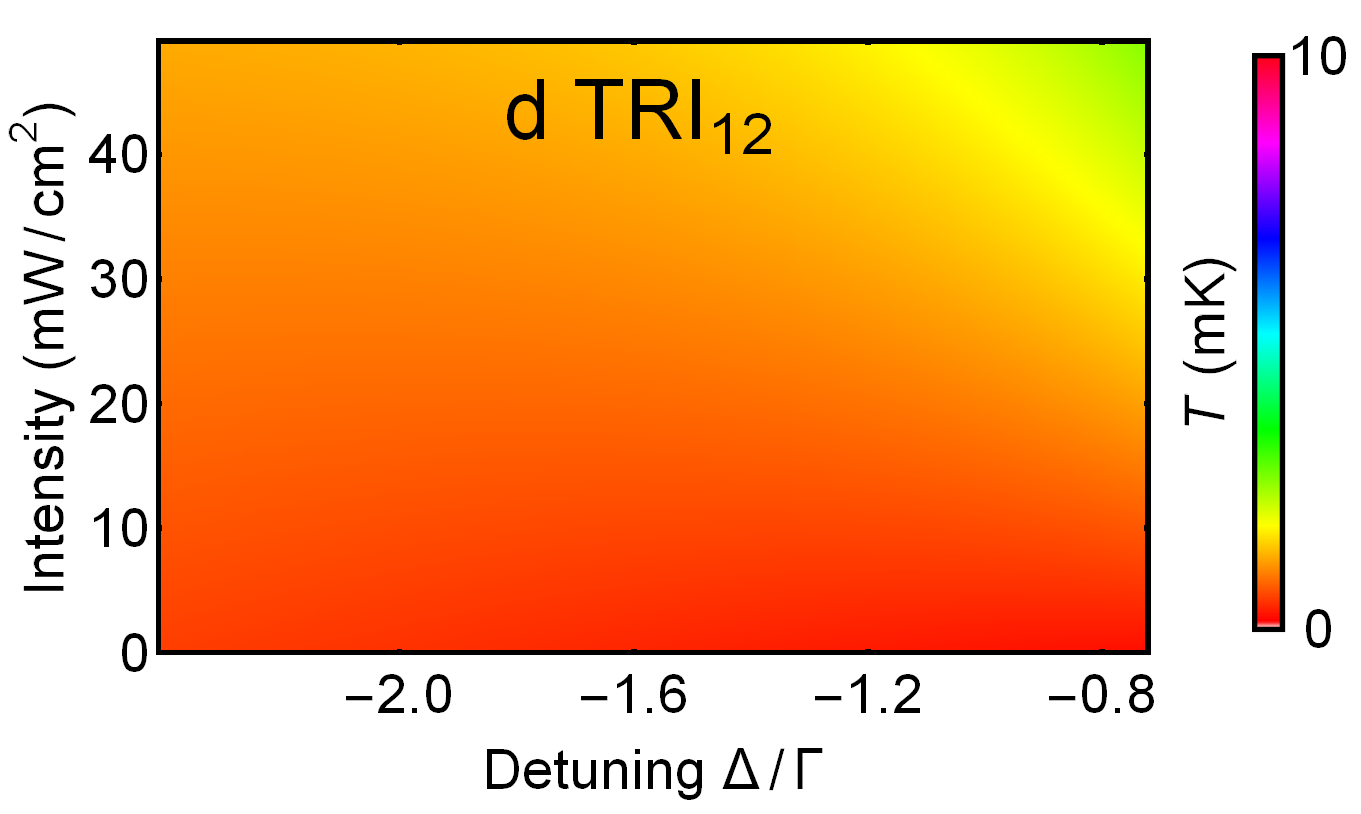}
\includegraphics[width=\columnwidth]{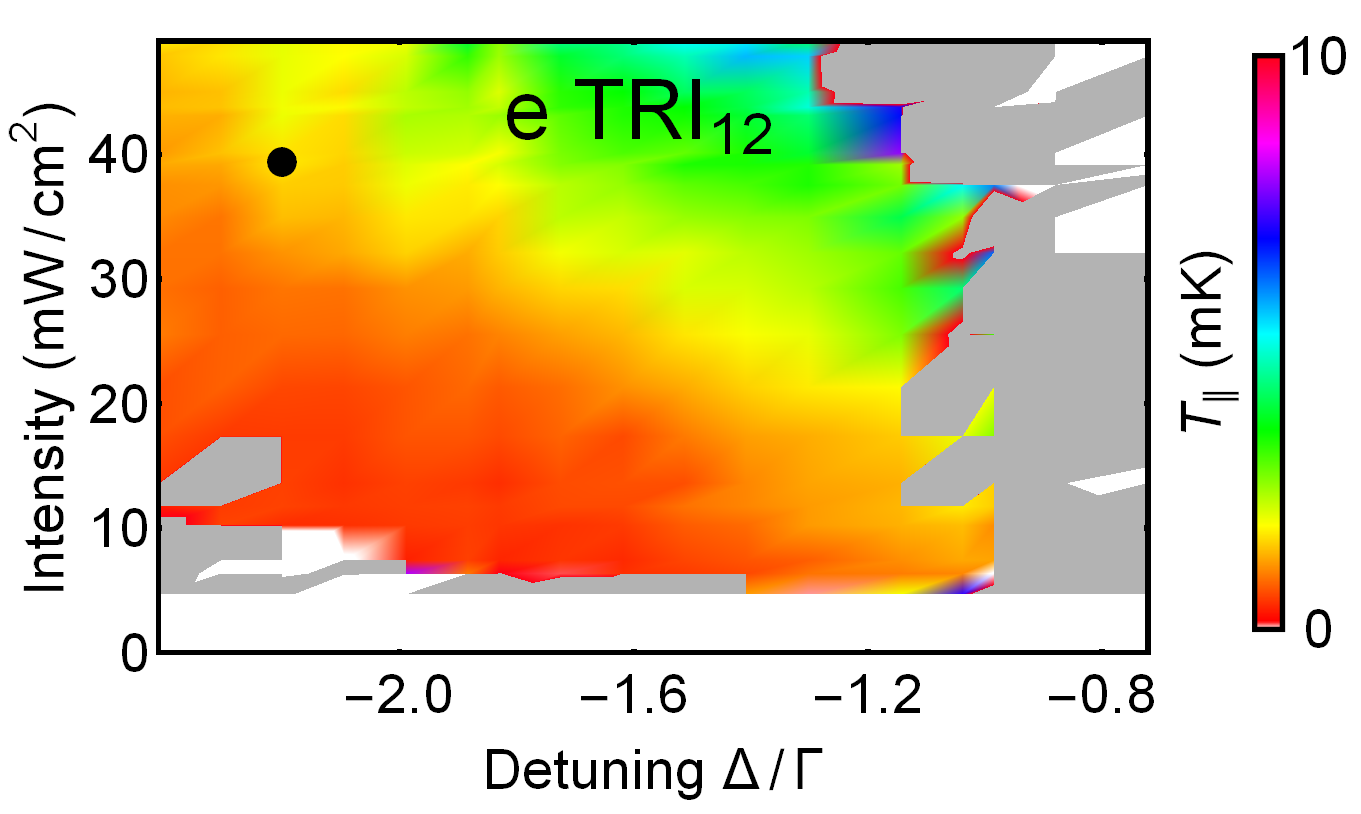}
\includegraphics[width=\columnwidth]{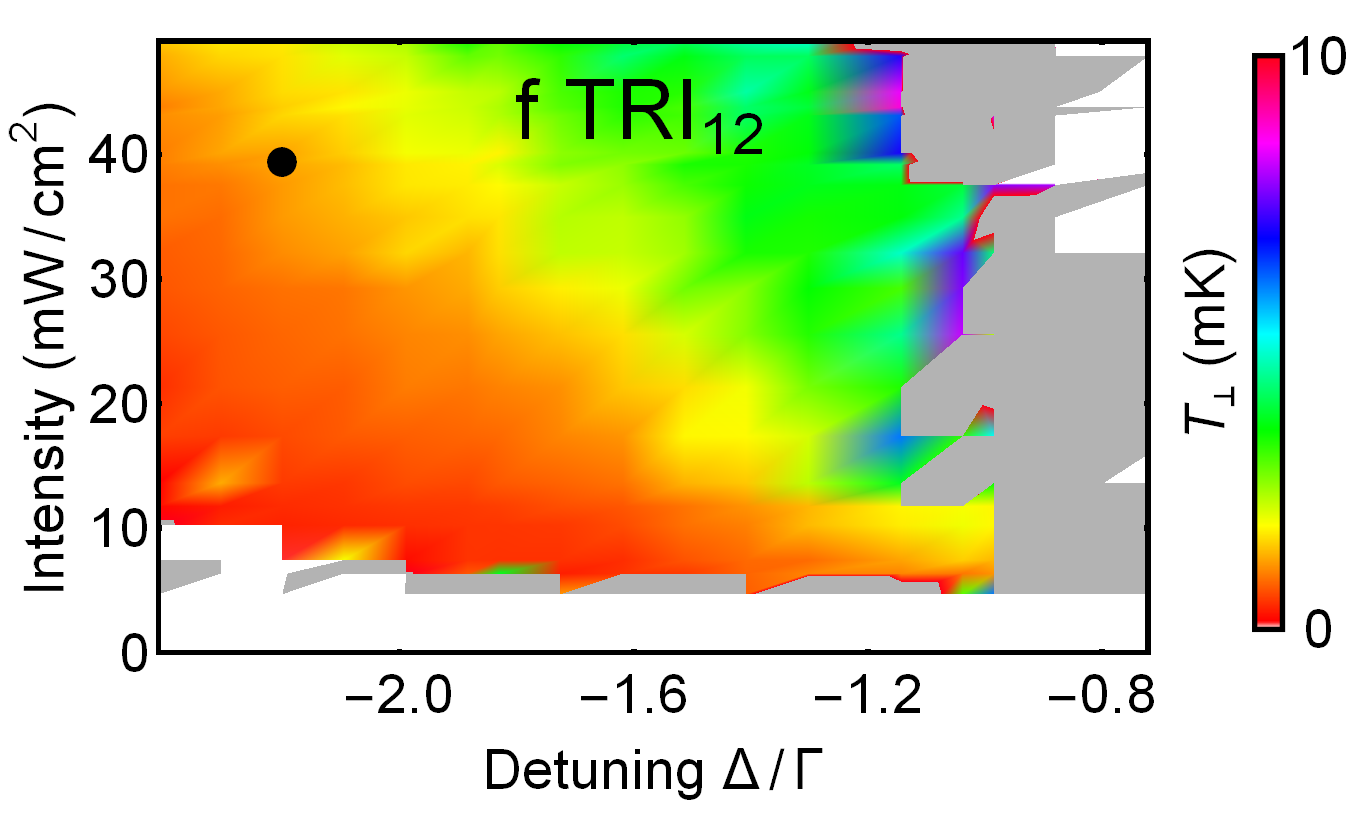}\end{minipage}
\caption{The full GMOT characterisation when using chips TRI$_{15}$ and TRI$_{12}$ with beam diffraction angles $\alpha=32^{\circ}$ and $41^{\circ}$, respectively. The labels a-f indicate theoretical atom number (a), experimental atom number (b) and spatial density (c), the theoretical temperature (d) and experimental temperatures in directions parallel (e) and perpendicular (f) to the grating. Light gray indicates indeterminate or out-of-range data. \href{http://photonics.phys.strath.ac.uk/?attachment_id=3507}{Media1} shows the theoretical temperature (d) for $I_S=1.67\,$mW/cm$^2$, which has clearer experimental agreement.\label{figtri}}
\end{figure}

\begin{figure}[!p]
\centering
\begin{minipage}{.3835\columnwidth}
\includegraphics[width=\columnwidth]{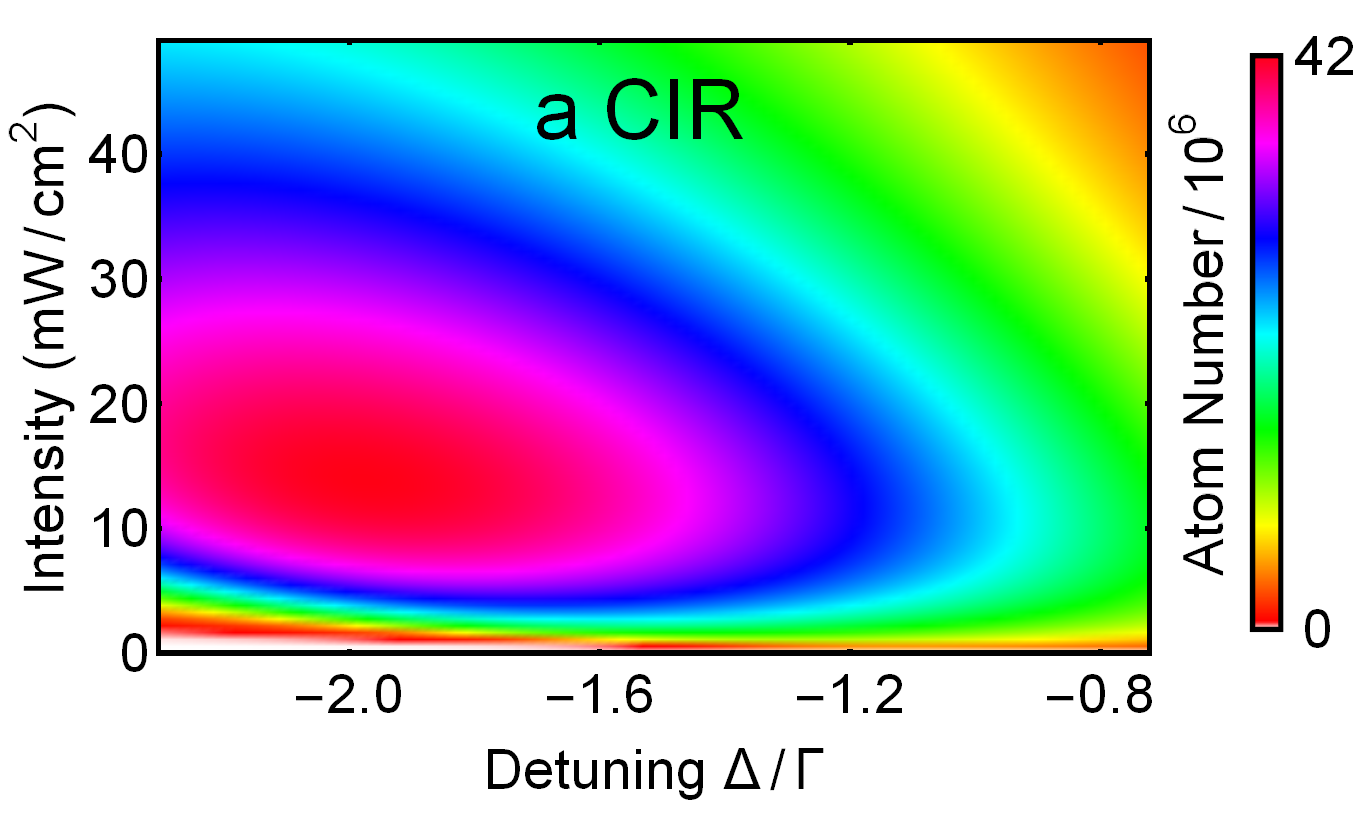}
\includegraphics[width=\columnwidth]{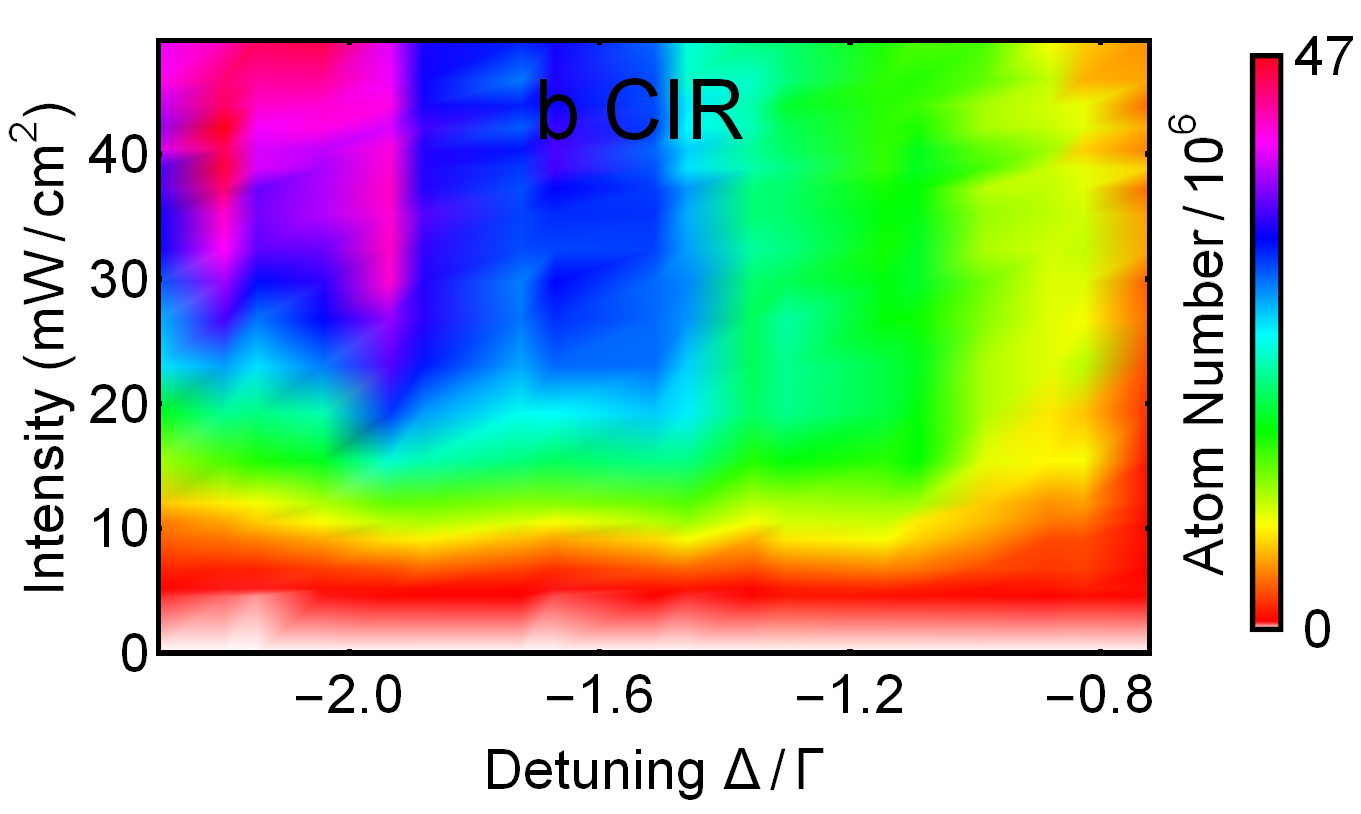}
\includegraphics[width=\columnwidth]{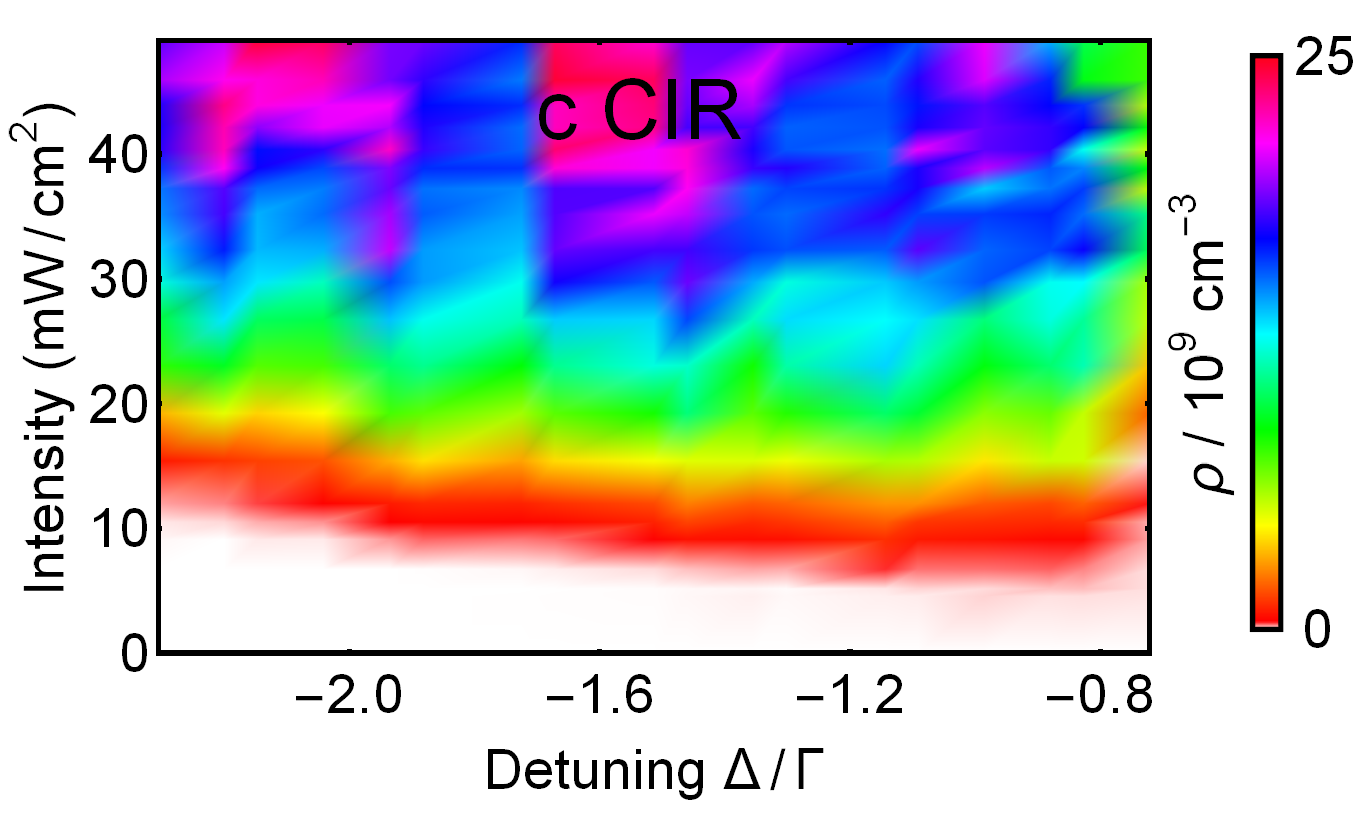}
\includegraphics[width=\columnwidth]{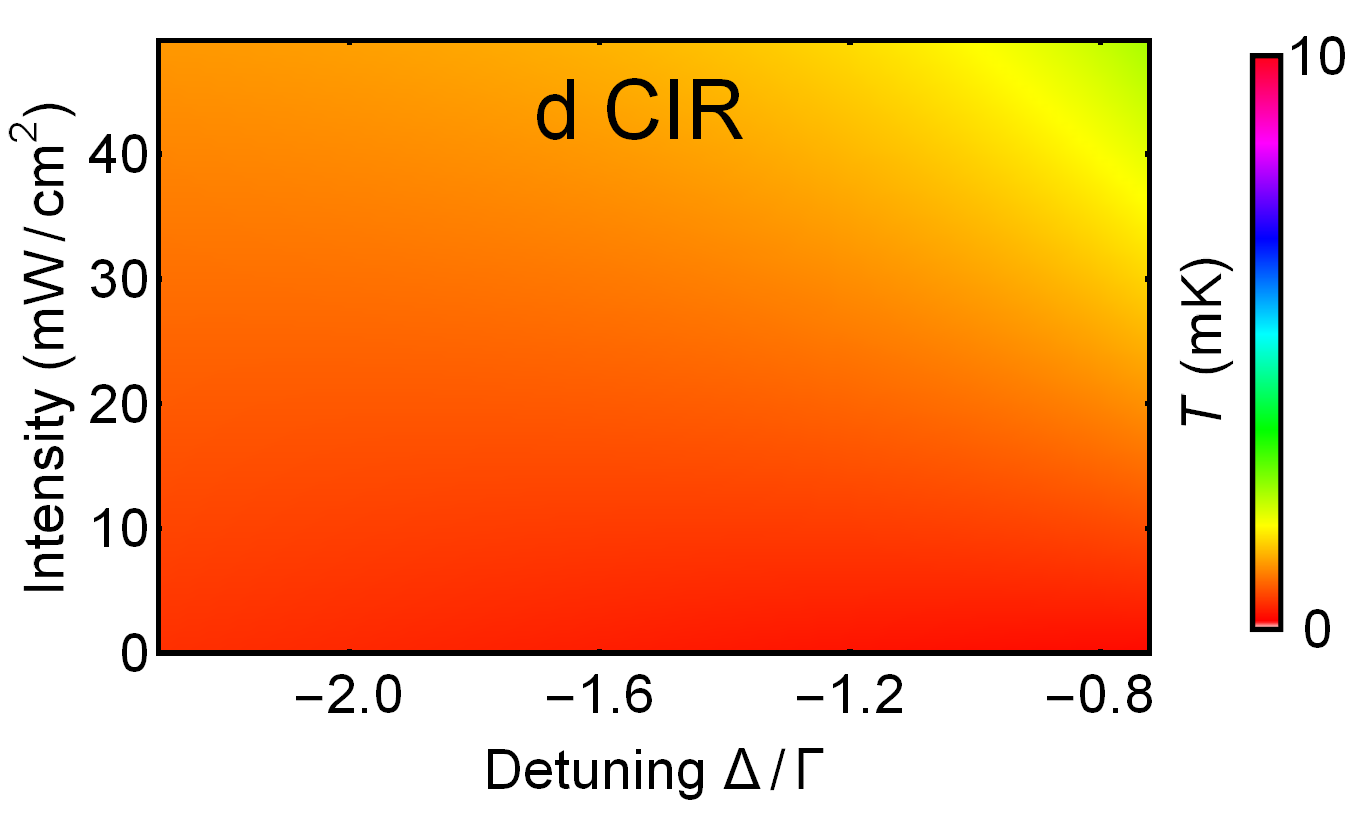}
\includegraphics[width=\columnwidth]{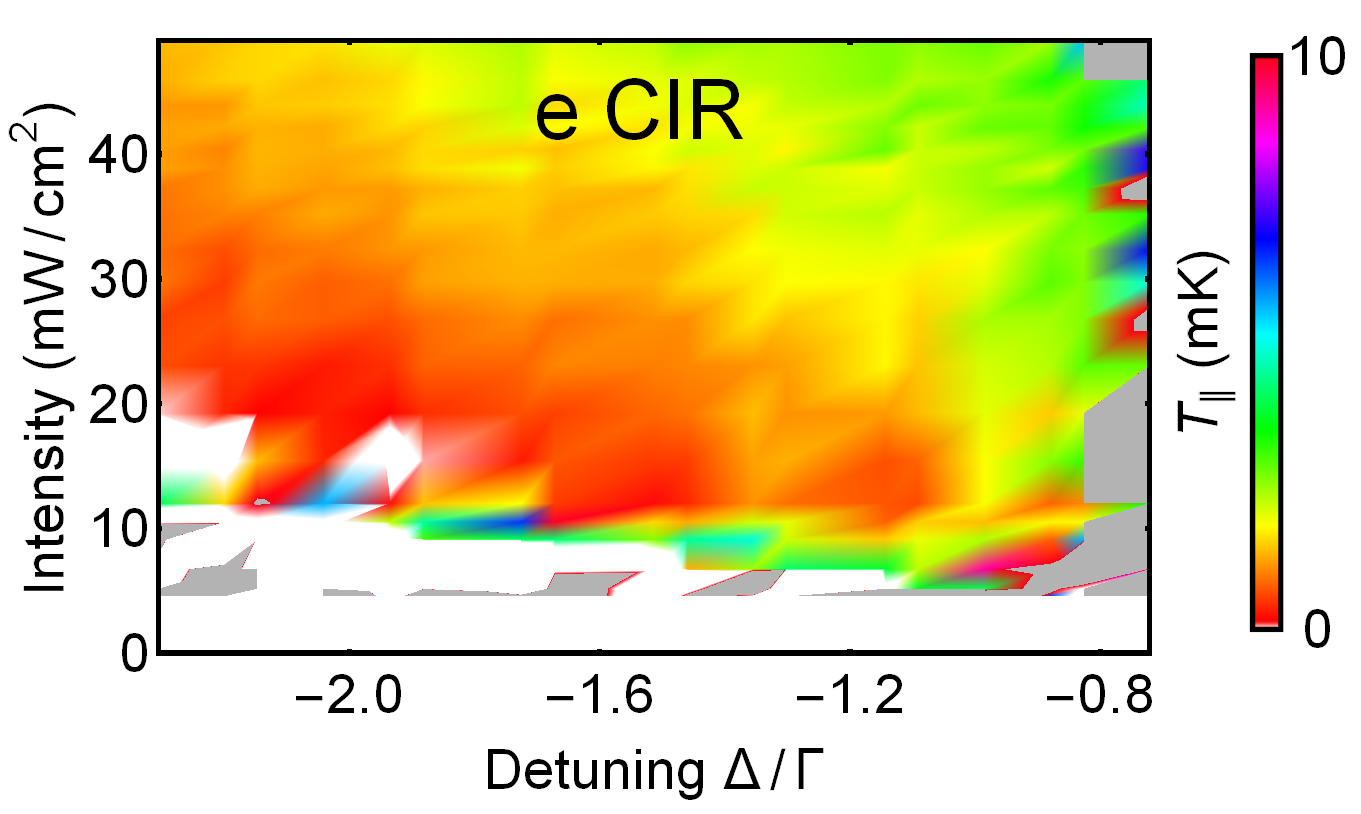}
\includegraphics[width=\columnwidth]{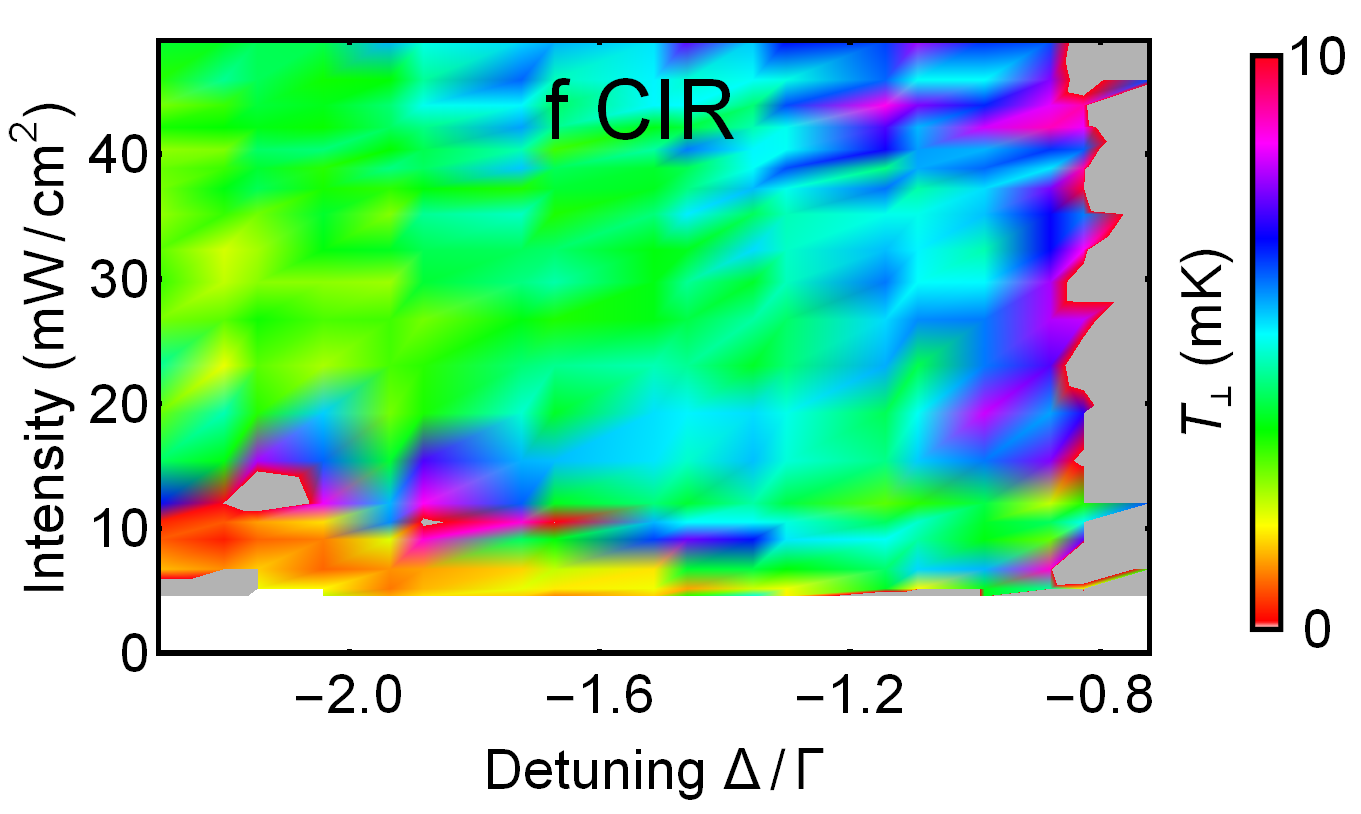}
\end{minipage}
\begin{minipage}{.3835\columnwidth}
\includegraphics[width=\columnwidth]{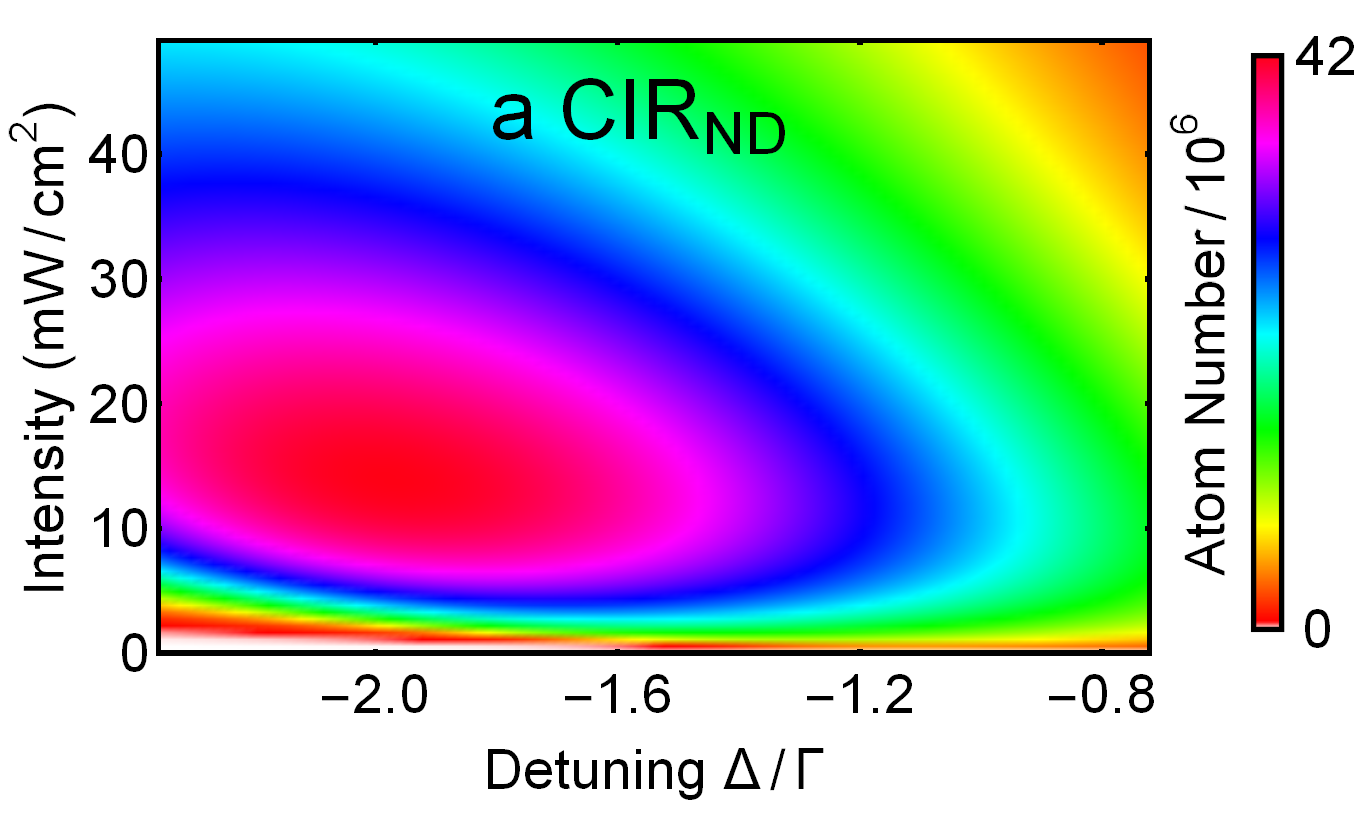}
\includegraphics[width=\columnwidth]{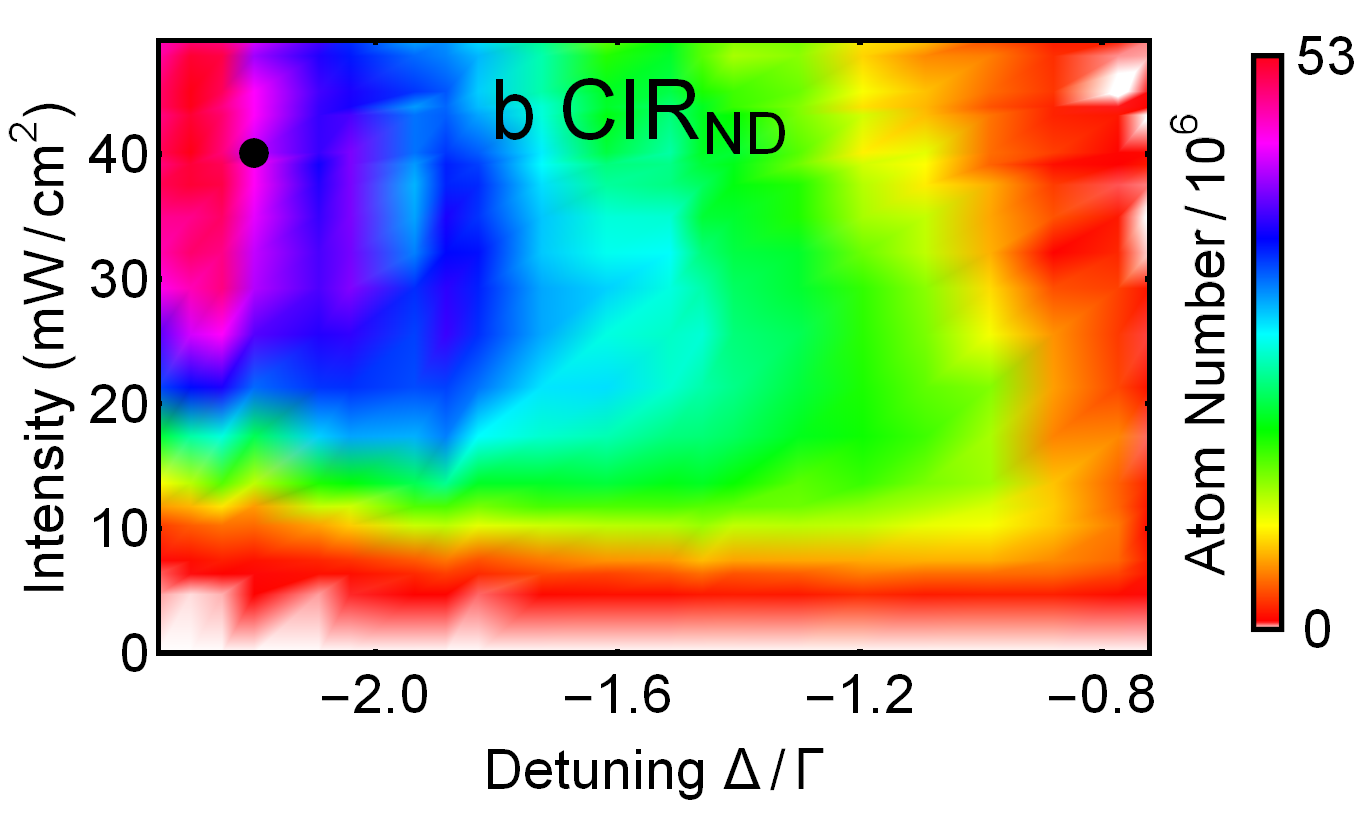}
\includegraphics[width=\columnwidth]{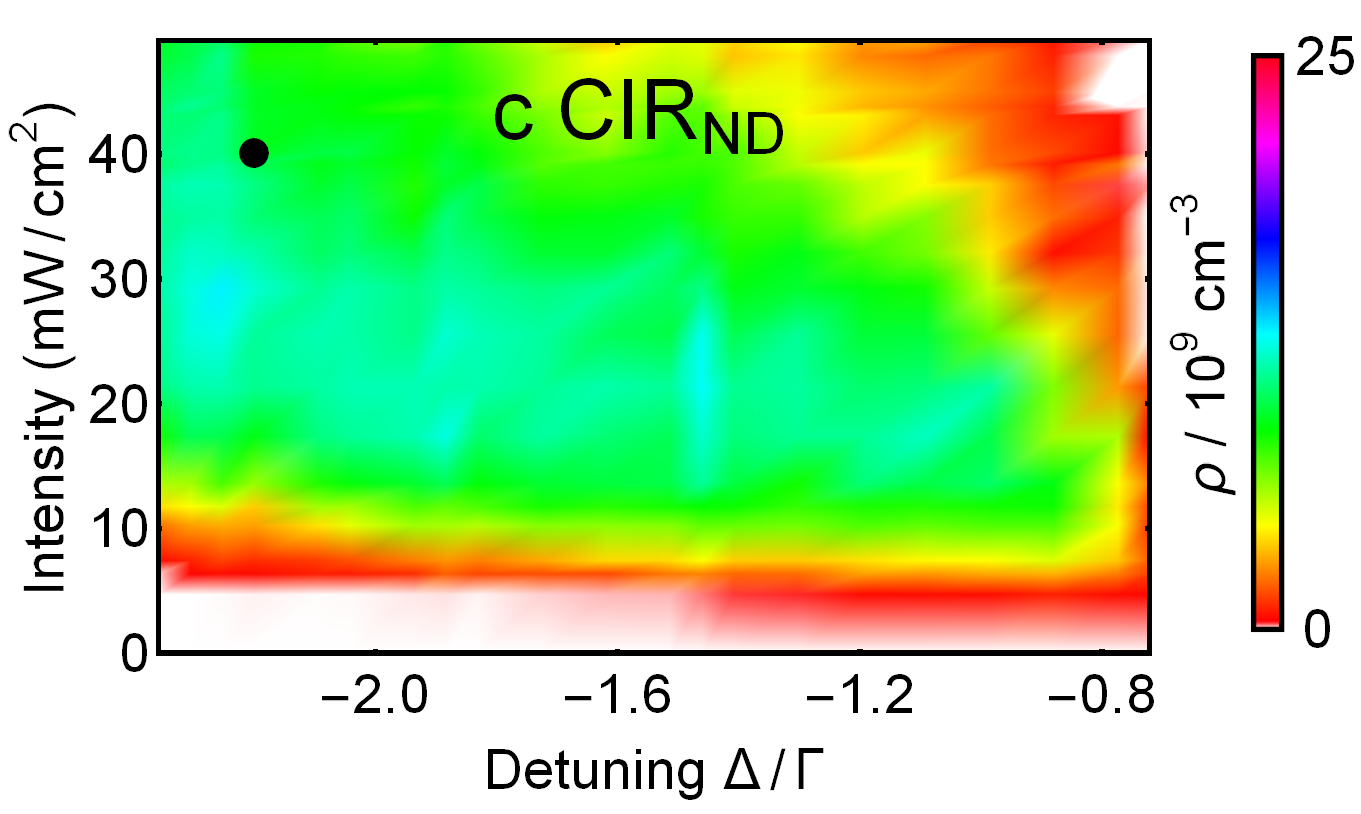}
\includegraphics[width=\columnwidth]{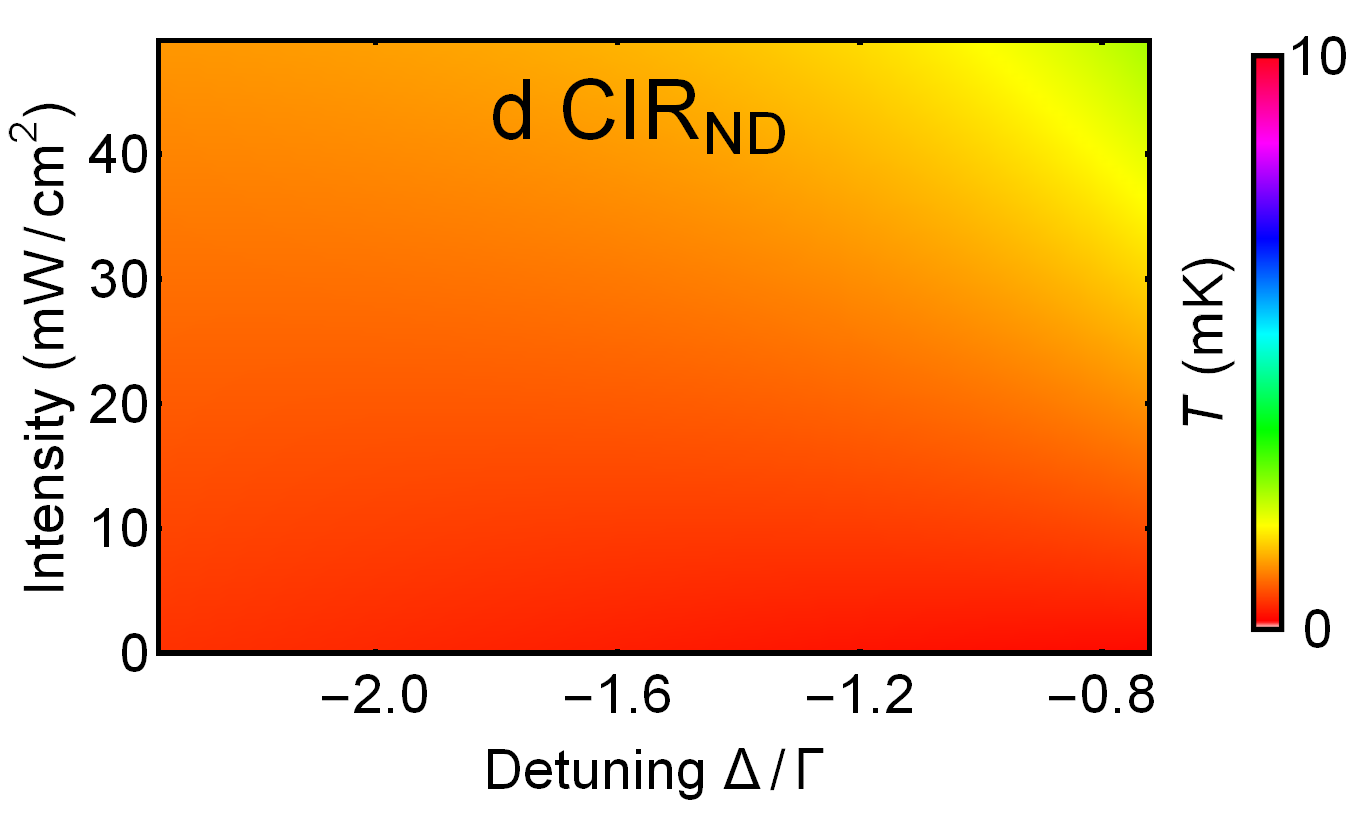}
\includegraphics[width=\columnwidth]{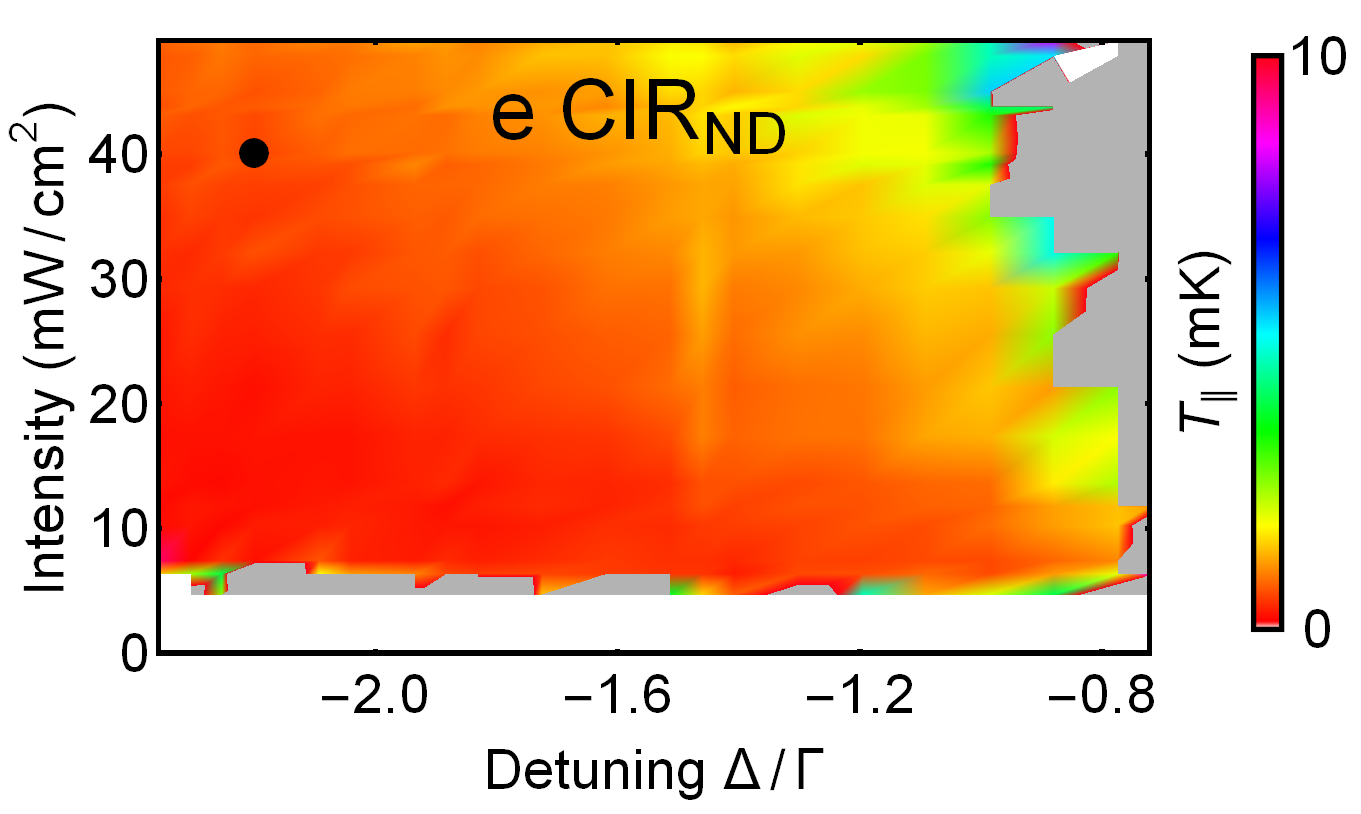}
\includegraphics[width=\columnwidth]{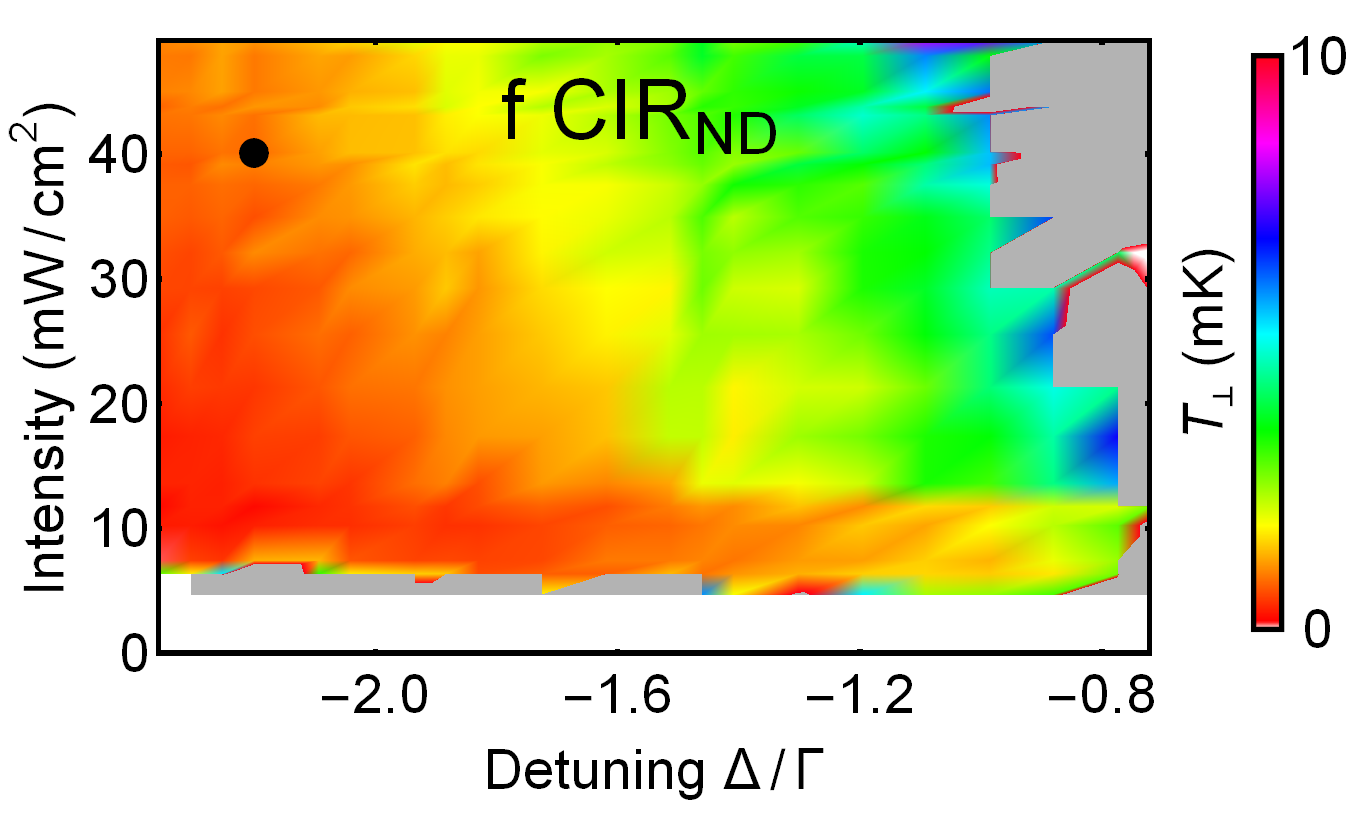}\end{minipage}
\caption{The full GMOT detuning and intensity characterisation when using chip CIR (standard Gaussian input beam a-f) and CIR$_\textrm{ND}$ (CIR with a $T=0.74$ ND filter in the central $3\,$mm diameter of the input beam, a1-f1), both with beam diffraction angle $\alpha=46^{\circ}$. Note the slight increase in atom number, and significant decrease in temperature using chip CIR$_\textrm{ND}$ relative to chip CIR. \href{http://photonics.phys.strath.ac.uk/?attachment_id=3508 }{Media2} is the $I_S=1.67\,$mW/cm$^2$ version of the theoretical temperature (images (d)). \label{figcir}}
\end{figure}

Figures \ref{figtri} (Chips TRI$_{15}$ and TRI$_{12}$) and \ref{figcir} (Chips CIR and CIR$_\textrm{ND}$) illustrate the  intensity- and detuning-dependence of a variety of MOT parameters: theoretical atom number (a), experimental MOT atomic number (b) and spatial density (c) as well as the theoretical Doppler temperature (d) with the temperatures experimentally measured in the directions both parallel (e) and perpendicular (f) to the grating. Note that the theoretical atom number saturates at about three times lower intensity than that observed experimentally -- we seem to need three times more laser power experimentally than we would theoretically expect. This discrepancy may be considered relatively minor given the very simple nature of the theoretical model, however we believe further investigation, and full comparison to a standard 6-beam MOT may still be warranted.

In prior measurements of $T_{\perp}$ on Chip CIR (Fig.~\ref{figcir}(f) CIR) there was an even larger discrepancy with theory (very high experimental temperatures) due to imbalanced optical force during the fluorescence imaging. This was determined via the noticeable centre-of-mass velocity accrued in the direction perpendicular to the grating during time-of-flight. In all the new data for Figs.~\ref{figtri} and \ref{figcir} any imbalanced radiation pressure (mainly for Fig.~\ref{figcir}) was balanced using a non-zero magnetic field oriented perpendicular to the grating. 

Note the significant increase in atom number ($\times 3$) and density ($\times 6$), with lower atomic temperature on chip TRI$_{12}$ compared to chip TRI$_{15}$ -- in future we intend to test gratings with even smaller period (larger diffraction angle) to determine if this trend continues.

\section{Experiment: best of both worlds}
\label{bbw}

For many metrological experiments, such as atomic clock measurements, the precision achieved strongly depends upon the temperature and total population of the atomic ensemble. Hence, an understanding of the final number of atoms that can be brought from the MOT stage to ultracold temperatures provides a good indication of the full capability of the GMOT. In Nshii \textit{et al.} \cite{nshii13} we effectively had only binary control of the intensity -- this meant we could \textit{either} get a large number of atoms on Chip CIR with high intensity, or good optical molasses on chip TRI$_{15}$ with lower intensity light (and a correspondingly small number of atoms loaded in the MOT).

After analysis of the results presented in Section \ref{exp}, the most suitable gratings for MOT number and temperature are Chip CIR$_\textrm{ND}$ and TRI$_{12}$. To study the number of atoms that could be brought to sub-Doppler temperatures, a MOT of intensity $40\,$mW/cm$^2$ and detuning of $\Delta/\Gamma = -2.2$ (initial MOT indicated by black dots in experimental Figs.~\ref{figtri} TRI$_{12}$ and \ref{figcir} CIR$_\textrm{ND}$) was transferred into optical molasses. The molasses consisted of turning off trapping magnetic fields, whilst using a two-step frequency jump of $\Delta/\Gamma=-5.3$ and $\Delta/\Gamma=-8.3$, each of duration $12\,$ms. 

\begin{figure}[!t]
\centering
\begin{minipage}{.49\columnwidth}
\includegraphics[width=\columnwidth]{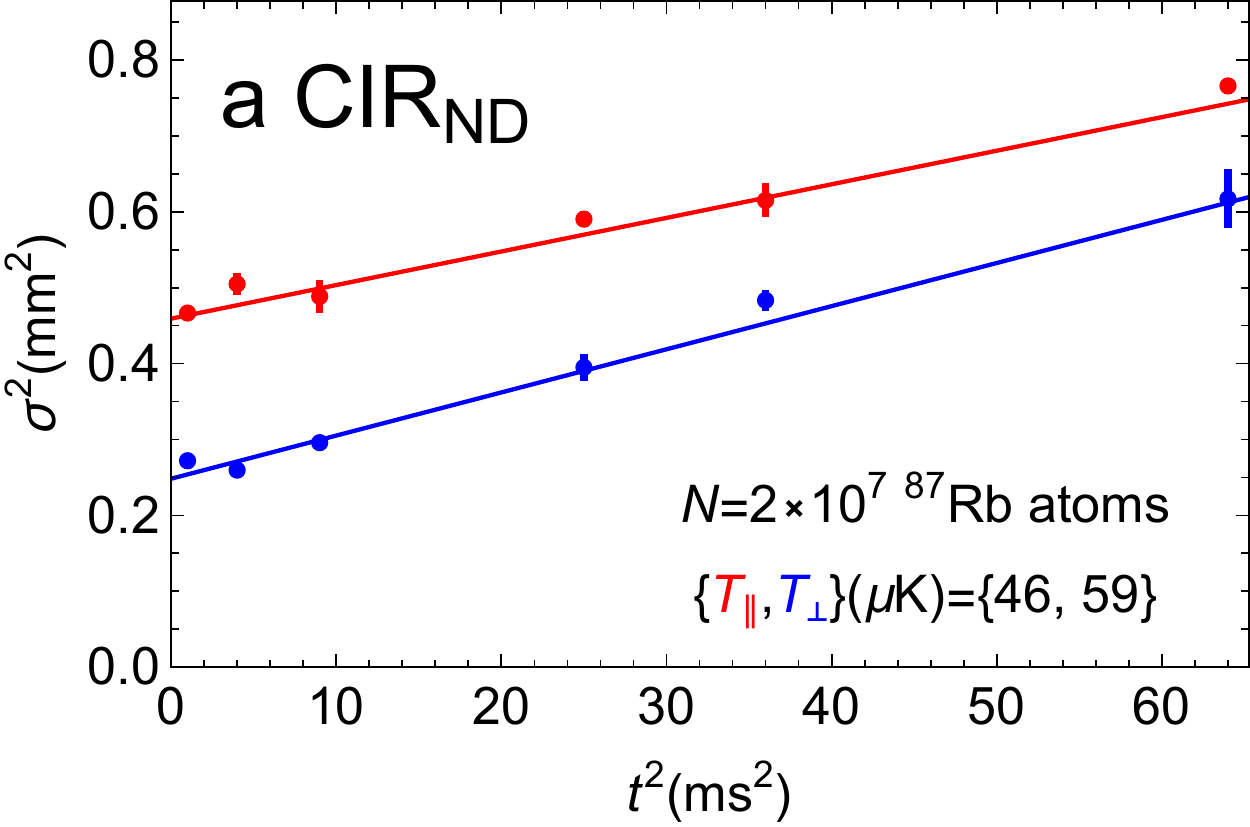}
\end{minipage}
\begin{minipage}{.49\columnwidth}
\includegraphics[width=\columnwidth]{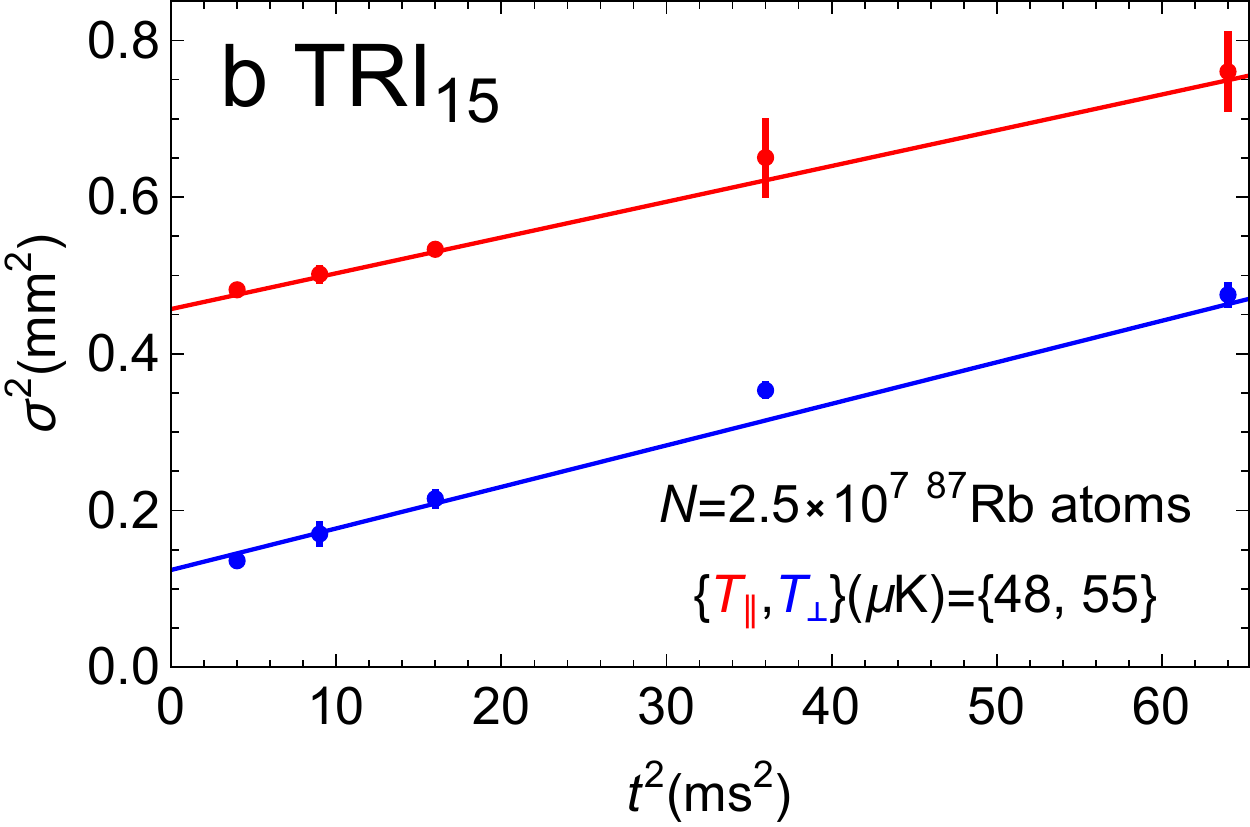}
\end{minipage}
\caption{Temperature measurements in directions both parallel (red) and perpendicular (blue) to the grating for chip TRI$_{12}$ (a) and CIR$_\textrm{ND}$ (b). Both chips give a 3D average temperature $T=\frac{2}{3}T_{\parallel}+\frac{1}{3}T_{\perp}=50\,\mu$K. \label{molass}}
\end{figure}

Under these conditions, Chip CIR$_\textrm{ND}$ brought $2\times 10^7$ atoms down to a 3D average temperature of 50~$\mu$K (Fig.~\ref{molass}(a)). Chip TRI$_{12}$ used the same settings with the exception that the molasses intensity was reduced to $23\,$mW/cm$^2$ -- bringing 2.5$\times 10^7$ atoms down to a 3D average temperature of $50\,\mu$K (Fig.~\ref{molass}(b)). Both temperatures are similar to those achieved by Nshii \textit{et al.} in \cite{nshii13} (with TRI$_{15}$), but with an order of magnitude more atoms after molasses ($N\approx2\times 10^6$ in \cite{nshii13}, due to the lack of intensity control discussed above). Temperature $T$ is determined by fitting $\sigma^2={\sigma_0}^2+k_B T t^2/m$ to the data, where $\sigma$ is the standard deviation of the Gaussian fit to the atomic cloud spatial distribution, $m$ is the mass of an $^{87}$Rb atom, $k_B$ is Boltzmann's constant and $\sigma_0$ is the initial $(t=0)$ standard deviation of the cloud.

\section{Interpretation and conclusions}

Both the experimental MOT atom number and cloud temperature demonstrate surprisingly good agreement with theory, given the simplicity of the theoretical models employed.

We note that higher densities are achieved on Chip CIR and CIR$_\textrm{ND}$, largely due to the larger diffraction angle $\alpha$, with higher associated damping $\gamma$ and trapping $\kappa$ coefficients. There is a dramatic density improvement with TRI$_\textrm{12}$ over TRI$_\textrm{15}$  and this trend may continue with higher values of $\alpha$ toward the tetrahedral configuration $\alpha=\arccos(1/3)$, with the caveat that the capture volume and grating diffraction efficiency will decrease. In Doppler theory the cloud size scales as $\sigma\propto\sqrt{\gamma T/\kappa},$ however we note from the relatively constant experimental MOT density for a variety of detunings and intensities that we are reaching sufficient atom numbers to likely be entering the MOT `constant density regime' where reradiation forces become more important than restoring forces \cite{rerad,rerad2,rerad3,rerad4}.

In terms of the temperature, significant sub-Doppler effects are not present, however we note that one would only expect sub-Doppler cooling under conditions of low intensity and large negative detunings, where we have insufficient data coverage to draw any further conclusions. 
The atom number is sufficiently large that effects where the MOT temperature is expected to increase with atom number $(T\propto N^{1/3})$ might also become evident \cite{TNlink,TNlink2}.

Compact quantum measurement devices are a burgeoning area \cite{tino,kitching,hims}, with grating technologies providing a new way to optically simplify transportable ultracold atom experiments. In future we hope to investigate the effect of grating diffraction angle $\alpha$ on both grating optical properties \cite{upcoming} and cooling performance. The current work could also be extended from investigating MOTs into sub-Doppler optical molasses regimes \cite{orozco}, preparing the path for future studies of on-chip Bose-Einstein condensates and extremely stable optical lattices \cite{nshii13}.

\section*{Acknowledgments}
We gratefully acknowledge support from ESA under contract 4000110231/13/NL/PA and the UK EPSRC. PFG
acknowledges the generous support of the Royal Society of Edinburgh.

\end{document}